\documentclass[aps,nofootinbib,notitlepage,longbibliography,twocolumn, superscriptaddress]{revtex4-1}
\usepackage{amsmath,amssymb}
\usepackage{slashed}
\usepackage{graphicx}
\usepackage{bm}
\usepackage{float}
\usepackage[T1]{fontenc}
\usepackage[utf8]{inputenc}
\usepackage{gauss} 
\usepackage[top=1.0in,bottom=1.0in,left=1.0in,right=1.0in]{geometry}
\usepackage[colorlinks,linkcolor=blue,citecolor=blue]{hyperref}
\usepackage{subfig}
\usepackage{xcolor}

\newcommand{\be}{\begin{equation}}
\newcommand{\ee}{\end{equation}}
\newcommand{\bea}{\begin{eqnarray}}
\newcommand{\eea}{\end{eqnarray}}
\newcommand{\ba}{\begin{eqnarray}}
\newcommand{\ea}{\end{eqnarray}}

\allowdisplaybreaks

\def\be{\begin{eqnarray}}
\def\ee{\end{eqnarray}}
\def\bea{\be}
\def\eea{\ee}

\def\roughly#1{\mathrel{\raise.3ex\hbox{$#1$\kern-.75em%
\lower1ex\hbox{$\sim$}}}}

\usepackage{lipsum}
\usepackage{soul}

\date{\today}
\begin{abstract}
We evaluate the leading exchange corrections to the Helium-4 
gravitational form factors (GFFs)  upto momenta of the order of the nucleon mass. 
We  use both  the K-harmonic method with simple pair nucleon potential,  and a Jastrow trial function using the Argonne $v_{14}$ potential, to evaluate the  Helium-4 GFFs. The exchange current contributions include the pair interaction, plus the seagull and the pion exchange interactions, modulo the recoil corrections.
To estimate the off-shellness of the pion nucleon coupling in this momenta range, we discuss the results using either the pseudo-scalar (PS) or pseudo-vector (PV) pion-nucleon couplings. When the PV coupling is used, the pair diagram contribution is higher order in the non-relativistic expansion. The results for the Helium-4 A-GFF  are comparable to those given by the impulse approximation, especially for the PS coupling using both the K-Harmonic method and variational method. The exchange current contributions with the PS coupling for the charge form factor of Helium-4, yield better agreement with the existing data over a broad range of momenta, especially when the Argonne $v_{14}$ potential including the D-wave admixture is used.
\end{abstract}
\begin{document}
\title{Helium-4 gravitational form factors: \\
exchange currents
}
\author{Fangcheng He}
\email{fangcheng.he@stonybrook.edu}
\affiliation{Center for Nuclear Theory, Department of Physics and Astronomy,
Stony Brook University, Stony Brook, New York 11794‚Äì3800, USA}

\author{Ismail Zahed}
\email{ismail.zahed@stonybrook.edu}
\affiliation{Center for Nuclear Theory, Department of Physics and Astronomy,
Stony Brook University, Stony Brook, New York 11794‚Äì3800, USA}
\maketitle

\section{Introduction}
The gravitational form factors (GFFs) of hadrons carry important information on their mass and spin structure.
Recent lattice QCD simulations of the pion and nucleon
GFFs have started to show how the quarks and gluons compose these form factors~\cite{Hackett:2023nkr,Hackett:2023rif,Wang:2024lrm}. This has led to 
a better understanding of the internal properties of
these hadrons~\cite{Polyakov:2018zvc,Burkert:2023wzr}, beyond their standard electromagnetic properties~\cite{Punjabi:2015bba}.

Recent measurements of threshold photoproduction of charmonium off nucleons by two collaborations at
JLab~\cite{GlueX:2019mkq,Duran:2022xag}, have given a first empirical estimate
of nucleon gluonic GFFs, with the help of model analyses~\cite{Mamo:2019mka,Guo:2021ibg}.  These measurements complement 
the  quark GFFs extracted from deeply virtual Compton scattering also at Jlab~\cite{CLAS:2015uuo}.  The combined empirical 
and numerical estimates of the nucleon GFFs, allow for
a better understanding of the mass and stress distribution inside the
nucleon~\cite{Abidin:2009hr,Belitsky:2002jp,Polyakov:2019lbq,Won:2023cyd,Hatta:2023fqc,Lorce:2024ipy}.  In particular, the mass radius of the 
nucleon was found to be smaller than the electromagnetic radius,
as suggested initially in dual gravity~\cite{Abidin:2009hr,Mamo:2022eui}.
Similar understanding is being also reached for the pion~\cite{Broniowski_2008,Frederico_2009,Fanelli_2016,Freese_2019,Krutov_2021,Raya:2021zrz,Xu:2023izo,Li_2024,Broniowski:2024oyk,Liu:2024,Liu:2024jno}.

Nuclei are composed of nucleons bound by strong interactions. Most of our understanding of nuclei
follows from  decades of electromagnetic scattering~\cite{Sick:2008zza},
showing that the electric and magnetic form factors receive sizable contributions from mesonic exchange currents in a wide range of momenta~\cite{Riska:1989bh}. Furthermore, electron and muon scattering on nuclei have also shown that the quark sub-constituents of the nucleon in nuclei get modified through shadowing (large Bjorken-x) and anti-shadowing (small Bjorken-x). However, it is less  clear how the gluon sub-constituents  are being affected 
in nuclei. Coherent photoproduction of phi-mesons off the
deuteron by the CLAS collaboration~\cite{CLAS:2007xhu} and LEPS collaborations~\cite{Chang:2007fc}, have not added more insights beyond those expected from two loosely bound nucleons.

Helium-4 is the lightest nucleus with strongly bound nucleons. It is a spin-isospin singlet with a well 
measured electromagnetic form factor, mass and charge
radius. Recently we have shown that the Helium-4 GFFs
in the impulse approximation, deviate from those of
independent nucleons in a range of momenta within the
nucleon mass~\cite{He:2023ogg}. Similar observations were made 
using the Skyrme model~\cite{GarciaMartin-Caro:2023klo}.
However, it is known that meson exchange-current contributions to the electromagnetic form factors are important~\cite{Hockert:1973fot,Chemtob:1974nf,Kloet:1973mj,Jackson:1975fys}.  The purpose of this work is to 
address these exchanges for the Helium-4 GFFs. For completeness, we note the light cone analysis of the deuteron GFFs in~\cite{Freese:2022yur}.

This paper is a follow up on our recent analysis of the EMT of light nuclei in the impulse approximation~\cite{He:2023ogg,He:2024vzz}. In section~\ref{SECII} we briefly detail the ground state wavefunctions for the description of Helium-4 using the K-harmonic method with no D-wave admixture, and the  variational method 
using the Argonne $v14$ potential with D-wave admixture. In section~\ref{SECIII} we describe the exchange current contributions 
to the GFFs of Helium-4, and their generic matrix elements with both 
wavefunctions. In section~\ref{SECIV} we explicitly evaluate all exchange matrix elements for Helium-4 using the wavefunction from the K-harmonic method. Those from the variational method are multi-dimensional and are evaluated numerically using Monte Carlo method. The numerical results are discussed in section~\ref{SECV}. Our conclusions are in section~\ref{SECVI}. In the Appendices we list the numerical results 
for the variational wavefunction for future use, and apply the same analysis for the Helium-4 charge form factor for comparison.

\section{Ground state of Helium-4}
\label{SECII}
In this section we briefly outline the construction of the ground state wavefunction of Helium-4, using two methods. The first method makes
use of the Reid nucleon pair potential and K-harmonics  in the hyper-radius approximation, without D-wave contribution. We recently used for the analysis of the GFFs of Helium-4 in the impulse approximation~\cite{He:2023ogg}. The second method, makes use of the Argonne $v14$ nucleon pair potentials,
retaining the D-wave contribution~\cite{Wiringa:1984tg}.

\label{APP_PIONX}
\subsection{K-Harmonic Method}
\begin{figure}
\begin{center}
\includegraphics[width=7cm, height=5.5cm]{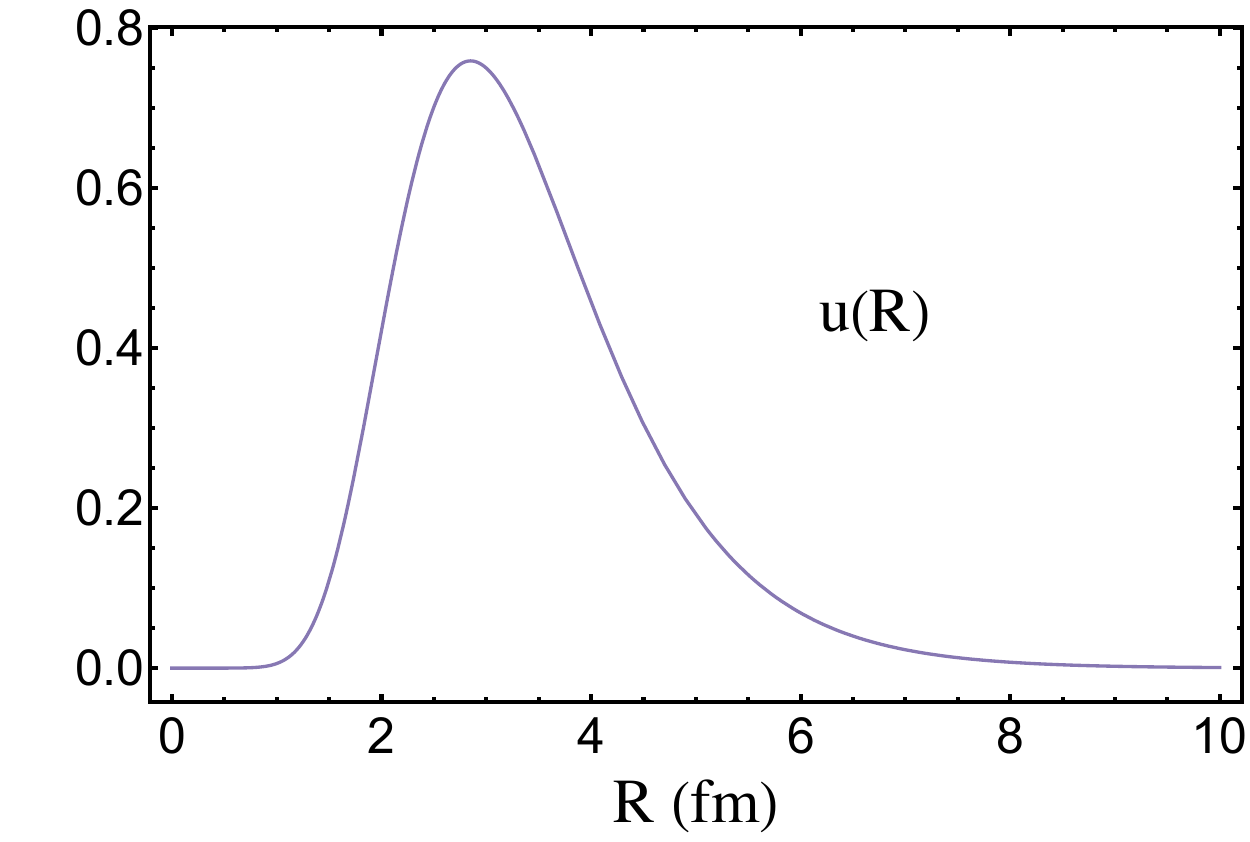}
\includegraphics[width=7cm, height=5.5cm]{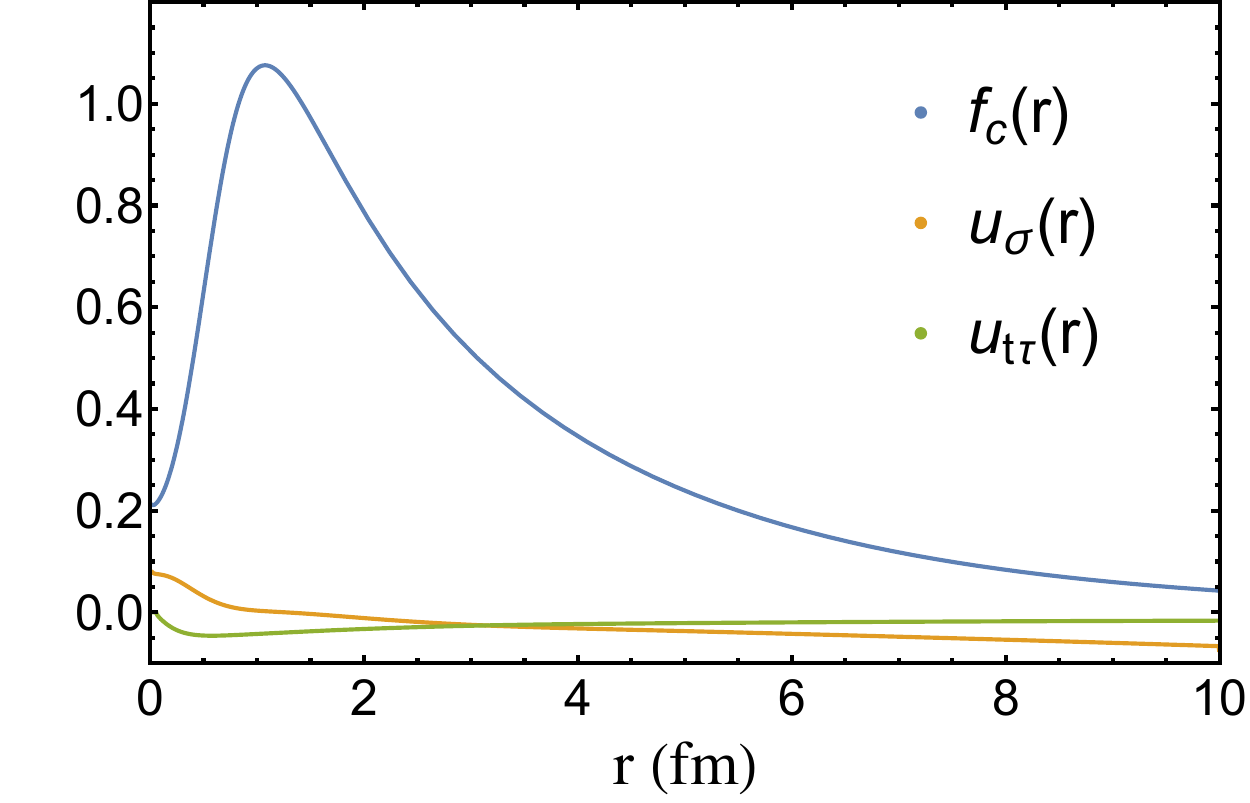}   
\caption{Top: The reduced wave function used in K-Harmonic Method; Bottom: Helium-4 parial wave-functions following from the use of the Argonne $v14$ potential~\cite{Wiringa:1991kp}:
central $f_c(r)$ (solid-bue), spin $u_\sigma(r)$ (solid-orange) and isospin-tensor $u_{t\tau}$ (solid-green).} 
\label{fig:wf_v14}
\end{center}
\end{figure}

 The essential aspects of the radial part of the ground state of Helium-4 using 
 the K-harmonic method in the hyper-radius approximation was discussed recently in~\cite{He:2023ogg} (and references therein). More specifically, the 4 nucleons in the $0^{++}$ spin-isospin
 configurations are described by 
\bea
\label{ALPHA1}
&&\Phi_H[1, ..., 4]=\varphi[r_1,r_2,r_3,r_4]\,\Phi[\sigma, \tau],
\eea
with $\Phi$ refers to the properly symmetrized spin-isospin wavefunction. To remove the spurious center of mass motion in Eq.~(\ref{ALPHA1}), the nucleon coordinates in~(\ref{ALPHA1}) are split into three Jacobi coordinates  $\vec{\xi}_{1,2,3}$ plus the center of mass $\vec R_C$
\bea\label{eq:Jacocorr}
\vec{\xi}_1&=&\frac{1}{\sqrt{2}}(\vec{r}_2-\vec{r}_1),\nonumber\\
\vec{\xi}_2&=&\frac{1}{\sqrt{6}}(\vec{r}_1+\vec{r}_2-2\vec{r}_3),\nonumber\\
\vec{\xi}_3&=&\frac{1}{2\sqrt{3}}(\vec{r}_1+\vec{r}_2+\vec{r}_3-3\vec{r}_4), \nonumber\\
\vec R_C&=&\frac 14 
(\vec r_1+\vec r_2+\vec r_3+\vec r_4),
\eea
with the radial hyperdistance
\bea
R^2= \frac 14 \sum_{i\neq j}
(\vec r_i-\vec r_j)^2=\vec \xi_1^2+\vec \xi_2^2+\vec \xi_3^2.
\eea
To keep this analysis simple as in~\cite{He:2023ogg}, we will ignore the 
D-wave contribution. Its role will be detailed in the second analysis to follow, using the Argonne potential below. 
With this in mind,  the radial eigenstates carry constant hyper-spherical harmonic, with the spin-isospin content~\cite{Castilho:1974}
\begin{widetext}
\bea
\label{eq:spispwf}
\Phi[\sigma,\tau] &=&\frac{\sqrt{105}}{8\pi^2}
\bigg(([\sigma(1),\sigma(2)]_1[\sigma(3),\sigma(4)]_1)_{00}
([\tau(1),\tau(2)]_0[\tau(3),\tau(4)]_0)_{00}\nonumber\\
&&\qquad\qquad -
([\sigma(1),\sigma(2)]_0[\sigma(3),\sigma(4)]_0)_{00}
([\tau(1),\tau(2)]_1[\tau(3),\tau(4)]_1)_{00}
\bigg).
\eea
\end{widetext}
The form of Eq.~(\ref{ALPHA1}) in hyper-spherical coordinates can be expressed as 
\bea
\Phi_H[1, ..., 4]=\varphi[R]\,\Phi[\sigma, \tau].
\eea
The radial part $\varphi[R]=u(R)/R^4$
where $u(R)$ is the reduced wavefunction, solves
\bea\label{eq:sch_Hel_new}
u{''}-\frac {12}{R^2} u-\frac{2m_N}{\hbar^2}
(W(R)+V_C(R)-E)u=0,\nonumber\\
\eea
Here $W(R)$ and $V_C$ represent the pair nucleon and Coulomb potentials, which can be written as~\cite{Castilho:1974,He:2023ogg}
\bea
\label{eq:WR}
W(R)&=&\frac{315}4\int_0^1\,dx\,(1-x^2)^2x^2\,
V(\sqrt 2R x),  \nonumber\\
V_C(R)&=&\frac{2.23{\rm MeV} {\rm fm}}{R},
\eea
with 
\bea\label{eq:V3}
V(r)&=&+1310.21\,e^{-(r/0.7)^2}-467.97\,e^{-(r/1.16)^2}.
\nonumber\\
\eea 
The radial solution to Eq.~(\ref{eq:sch_Hel_new}) is shown in the top panel in Fig.~\ref{fig:wf_v14}.
The ensuing wavefunction reproduces the correct binding energy and charge radius of Helium-4. It also describes fairly well
the empirical charge form factor of Helium-4 up-to momentum of the order of half the nucleon mass~\cite{He:2023ogg}.

\subsection{Argonne $v14$ variational method}
To include the D-wave contribution and more realistic nucleon pair interactions in Helium-4,
we will make use of the Argonne $v14$ symmetrized ($\cal S$) variational wavefunction~\cite{Lomnitz-Adler:1980rud, Lomnitz-Adler:1981dmh, Wiringa:1991kp}
\bea\label{eq:Jastraw}
|\Psi_v\rangle = {\cal S}\prod \limits_{i<j}(1+U_{ij})|\Psi_J\rangle,
\eea
with the Jastrow trial state
$$|\Psi_J\rangle=\prod \limits_{i<j} f_c(r_{ij})|\Phi\rangle,$$ 
where $f_c(r)$ is a central pair correlation function, and $\Phi$ 
the antisymmetric spin-isospin wavefunction in Eq.~(\ref{eq:spispwf}). 
The pair correlation functions $U_{ij}$ built in the variational state, is valued
in the set of operators making up the Argonne $v14$ potential
\bea
U_{ij}=\sum_{p=2}^mu_p(r_{ij})O_{ij}^p,
\eea
with $O_{ij}^{p}$ making up the 14 spin-isospin  pair operators in the $v14$ potential~\cite{Lomnitz-Adler:1980rud, Lomnitz-Adler:1981dmh, Wiringa:1991kp}
\bea
O_{ij}^{p=1,14}&=&1,\tau_i\cdot\tau_j,\sigma_i\cdot\sigma_j, (\tau_i\cdot\tau_j)(\sigma_i\cdot\sigma_j), S_{ij}, \nonumber\\
&&S_{ij}(\tau_i\tau_j), \vec{L}.\vec{S}, \vec{L}.\vec{S}(\tau_i\tau_j), \vec{L}^2, \vec{L}^2(\tau_i\cdot\tau_j),  \nonumber\\
&&\vec{L}^2(\sigma_i\cdot\sigma_j), \vec{L}^2(\sigma_i\cdot\sigma_j)(\tau_i\cdot\tau_j), \nonumber\\
&& (\vec{L}.\vec{S})^2, (\vec{L}.\vec{S})^2(\tau_i\tau_j).
\eea
Numerically the trial functions $u_p(r_{ij})$ will turn out to be small, which allows us to simplify Eq.~\eqref{eq:Jastraw}
\bea\label{eq:Jastrawsim}
|\Psi_v\rangle = (1+\sum_{i<j}U_{ij})|\Psi_J\rangle.
\eea
Using the spin-isospin identities~\cite{Lomnitz-Adler:1980rud}
\bea\label{eq:gammataurel}
\tau_i\cdot\tau_j|\Phi\rangle&=&-(2+\sigma_i\cdot\sigma_j)|\Phi\rangle, \nonumber\\
(\tau_i\cdot\tau_j)(\sigma_i\cdot\sigma_j)\Phi\rangle&=&-3|\Phi\rangle, \nonumber\\
S_{ij}(\tau_i\cdot\tau_j)\Phi\rangle&=&-3S_{ij}|\Phi\rangle, \nonumber\\
(\vec{L}.\vec{S})|\Phi\rangle&=&\vec{L}^2|\Phi\rangle=0.
\eea
To include the central three body correlation, the pair correlation function can be modified as follows~\cite{Lomnitz-Adler:1980rud, Lomnitz-Adler:1981dmh, Wiringa:1991kp}.
\bea
u_p(r_{ij})\rightarrow {\cal F}_{ij}u_p(r_{ij})=[\Pi_{k\neq i,j}f_{ijk}]u_p(r_{ij}),
\eea
where $f_{ijk}$ is chosen to be~\cite{Wiringa:1991kp}
\bea
f_{ijk}&=&1-9\left(\frac{r_{ij}}{r_{ij}+r_{ik}+r_{jk}}\right)^4  \nonumber\\
&\times&\textbf{exp}[-0.1(r_{ij}+r_{ik}+r_{jk})],
\eea
we can rewrite Eq.~\eqref{eq:Jastrawsim} in the form
\bea\label{eq:Jastrawsim1}
&&|\Psi_v\rangle =\Big\{1+\sum_{i<j}{\cal F}_{ij}\big[u_\sigma(r_{ij})\sigma_i.\sigma_j  \nonumber\\
&&+u_{t\tau}(r_{ij})S_{ij}\tau_i.\tau_j)\big]\Big\}|\Psi_J\rangle.
\eea
The Jastrow pair correlation function $f(r)$ and the operator correlation functions  $u_{\sigma, t\tau}$, can be repacked
in  Eq.~(\ref{eq:Jastrawsim1}), in terms of  their contributions to the  singlet and triplet spin and isospin channels, using the identity~\cite{Wiringa:1991kp,Lomnitz-Adler:1981dmh}
\bea
&&f_c(r_{ij})\left[1+u_\sigma(r_{ij})\sigma_i.\sigma_j+u_{t\tau}(r_{ij})\sigma_i.\sigma_j\right]|\Phi\rangle\nonumber\\
&&\hspace{-0.5cm}=\bigg(\sum_{S,T}f_{S,T}(r_{ij}) P_SP_T+f_{t,0}(r_{ij})S_{ij}P_{T=0}\bigg)|\Phi\rangle.
\eea
Here $P_S$ and $P_T$ are the spin and isospin projection operators, i.e. 
\bea
P_{S=1}=(3+\sigma_i.\sigma_j)/4,\nonumber\\ P_{S=0}=(1-\sigma_i.\sigma_j)/4, \nonumber\\
P_{T=1}=(3+\tau_i.\tau_j)/4,\nonumber\\ P_{T=0}=(1-\tau_i.\tau_j)/4.
\eea 
The pair correlation functions $f_c$, $u_\sigma$ and $u_{t\tau}$ can be related to the projected functions through
\bea
f_c&=&\frac{1}{4}(3f_{1,0}+f_{0,1}),\nonumber\\
u_\sigma&=&\frac{1}{4}(f_{1,0}-f_{0,1})/f_c, \nonumber\\
u_{t\tau}&=&-\frac{1}{3}f_{t,0}/f_c,
\eea
where we made use of Eq.~(\ref{eq:gammataurel}). In the ground state of Helium-4 with $L=0$,
the spin-isospin projected functions $f_{S,T}$ and spin-singlet-tensor function $f_{t,0}$ can be obtained by solving
the coupled Schrodinger equations~\cite{Wiringa:1991kp}
\begin{widetext}
\bea\label{eq:seq}
&&-\frac{\hbar^2}{m}[r\,f_{0,1}(r)]''+(\bar{v}_{0,0}+\lambda_{0,T})f_{0,1}(r)r=0,
\nonumber\\
&&-\frac{\hbar^2}{m}[r\,f_{1,0}(r)]''+(\bar{v}_{1,0}+\lambda_{1,0})f_{1,0}(r)r
+8(\bar{v}_{t,0}+\lambda_{t,0})f_{t,0}(r)r=0,
\nonumber\\
&&-\frac{\hbar^2}{m}[(r\,f_{t,0}(r))''-\frac{6}{r^2}(r\,f_{t,0}(r))]+(\bar{v}_{t,0}+\lambda_{t,0})f_{1,0}(r)r
\nonumber\\
&&~~~~~~+\left(\bar{v}_{t,0}+\lambda_{t,0}-2(\bar{v}_{t,0}+\lambda_{t,0})-3(\bar{v}_{b,0}+\lambda_{b,0})+6\bar{v}_{q1,0}+q\bar{v}_{bb,0}\right)f_{t,0}(r)r=0,
\eea
\end{widetext}
here  $\bar{v}_{S,T}$ refer to the explicit $v14$ pair nucleon potentials.
The eigenvalues $\lambda_{S,T}$ include the screening effects at short distances, and their behavior at large distances is fixed by the asymptotic of the correlation functions. We refer to~\cite{Wiringa:1984tg, Wiringa:1991kp} for their explicit form. In Fig~\ref{fig:wf_v14} we show our numerical results 
for $f_c$, $u_\sigma$ and $u_{t\tau}$, which are in overall agreement with those reported in~\cite{Wiringa:1991kp}. In  appendix.~\ref{sec:wfv14},
we have tabulated our results  for future comparison.

\section{Exchange currents in Helium-4}
\label{SECIII}
Helium-4 gravitational form factors in the impulse approximation were discussed  in~\cite{He:2023ogg}. In this section we extend the analysis to the
exchange currents, following our recent discussion in the deuteron case~\cite{He:2024vzz}. More specifically, the gravitational exchange contributions to Helium-4
in the single pion exchange approximation are illustrated in  Fig.~\ref{fig:NR_diagrams}. We will refer to them generically by $T_X^{\mu\nu}$ with the pair contribution (a), the seagull contribution (b), the
pion exchange contribution (c) and the recoil corrections plus wavefunction renormalization (d).  For a further assessment of the off-shellness effects between the pion-nucleon pseudoscalar and pseudovector couplings, we will discuss both, where we note that the seagull contribution (c) is absent for the former coupling. The explicit operator forms for each of these contributions are quoted in section~\ref{SECIV} below.

\subsection{Matrix element: K-harmonic method}
The generic structure of the matrix element in the K-harmonic method, can be recast into a multi-dimensional
integral over the Jacobi coordinates
\begin{widetext}
\bea
\label{PAIR0JHE4}
&&\left<+\frac{k}{2} m'\bigg|T_{X}^{\mu\nu}\bigg|-\frac{k}{2} m\right>
\nonumber\\
&=&\int d^3\xi_1\int d^3\xi_2\int d^3\xi_3\frac{d^3q_2}{(2\pi)^3}
\nonumber\\
&\times&\textbf{exp}\Bigg[-i\sqrt{2}\vec{q}_2\vec{\xi}_1+i\vec{k}\left(\frac{\vec{\xi}_1}{\sqrt{2}}-\frac{\vec{\xi}_2}{\sqrt{6}}-\frac{\vec{\xi}_3}{2\sqrt{3}}\right)\Bigg]
\langle\Phi_H[\xi]|T_{X}^{\mu\nu}(\sigma_1,\sigma_2)|\Phi_H[\xi]\rangle
+{\rm permutation}\nonumber\\
\eea
\end{widetext}
with  $\vec{q}_2$ the momentum of exchanged pion in Fig.~\ref{fig:NR_diagrams}. The argument of the Fourier transform
refers to the struck nucleon relative to the center of mass coordinate, i.e. 
\bea
\frac{\vec{\xi}_1}{\sqrt{2}}-\frac{\vec{\xi}_2}{\sqrt{6}}-\frac{\vec{\xi}_3}{2\sqrt{3}}=\vec{R}_C-\vec{r}_1
\eea
$T_{X}^{\mu\nu}$ is short for the inserted  EMT operators in Fig.~\ref{fig:NR_diagrams}.  Their explicit non-relativistic forms are listed in Eqs.~(\ref{eq:Tmunupair}), 
(\ref{eq:TmunuS}), (\ref{eq:TmunuPi}) and (\ref{eq:TmunuR}) for the pair contribution, the seagull contribution, the pion exchange contribution and recoil contribution, respectively. 
To keep the hyperspherical symmetric, the Jacobi coordinates are set to be 
\bea
\xi_1&=&R\,cos\theta(sin\theta_1cos\phi_1,sin\theta_1sin\phi_1,cos\theta_1),
\nonumber\\
\xi_2&=&R\,sin\theta cos\phi(sin\theta_2cos\phi_2,sin\theta_2sin\phi_2,cos\theta_2),
\nonumber\\
\xi_3&=&R\,sin\theta sin\phi(sin\theta_3cos\phi_3,sin\theta_3sin\phi_3,cos\theta_3),\nonumber\\
\eea
\begin{widetext}
The above integral can be reduced after integrating over the angular
\bea
\label{PAIR0JHE4_int}
&&\left<+\frac{k}{2} m'\bigg|T_{X}^{\mu\nu}\bigg|-\frac{k}{2} m\right>
\nonumber\\
&=&\int_0^\infty dR \int_0^\frac{\pi}{2} d\theta\,sin\theta^5 cos\theta^2\int_0^\frac{\pi}{2}d\phi\,sin\phi^2 cos\phi^2\int d\Omega_2\textbf{exp}\Bigg[-i\vec{k}\frac{\vec{\xi}_2}{\sqrt{6}}\Bigg]\int d\Omega_3\textbf{exp}\Bigg[-i\vec{k}\frac{\vec{\xi}_3}{2\sqrt{3}}\Bigg]
\nonumber\\
&\times&\int\frac{d^3q_2}{(2\pi)^3}\int d\Omega_1\textbf{exp}\Bigg[-i\sqrt{2}\vec{q}_2\vec{\xi}_1+i\vec{k}\frac{\vec{\xi}_1}{\sqrt{2}}\Bigg]u(R)^2\langle\Phi|T_{X}^{\mu\nu}(\sigma_1,\sigma_2)|\Phi\rangle+{\rm permutation}\nonumber\\
&=&(4\pi)^2\int_0^\infty dR \int_0^\frac{\pi}{2} d\theta\,sin\theta^5 cos\theta^2\int_0^\frac{\pi}{2}d\phi\,sin\phi^2 cos\phi^2 j_0\left(\frac{kR\,sin\theta cos\phi}{\sqrt{6}}\right)j_0\left(\frac{kR\,sin\theta sin\phi}{2\sqrt{3}}\right)
\nonumber\\
&\times&\int\frac{d^3q_2}{(2\pi)^3}\int d\Omega_1\textbf{exp}\Bigg[-i\sqrt{2}\vec{q}_2\vec{\xi}_1+i\vec{k}\frac{\vec{\xi}_1}{\sqrt{2}}\Bigg]u(R)^2\langle\Phi|T_{X}^{\mu\nu}(\sigma_1,\sigma_2)|\Phi\rangle+{\rm permutation},
\eea

\begin{figure*}
\begin{center}
\includegraphics[scale=0.6]{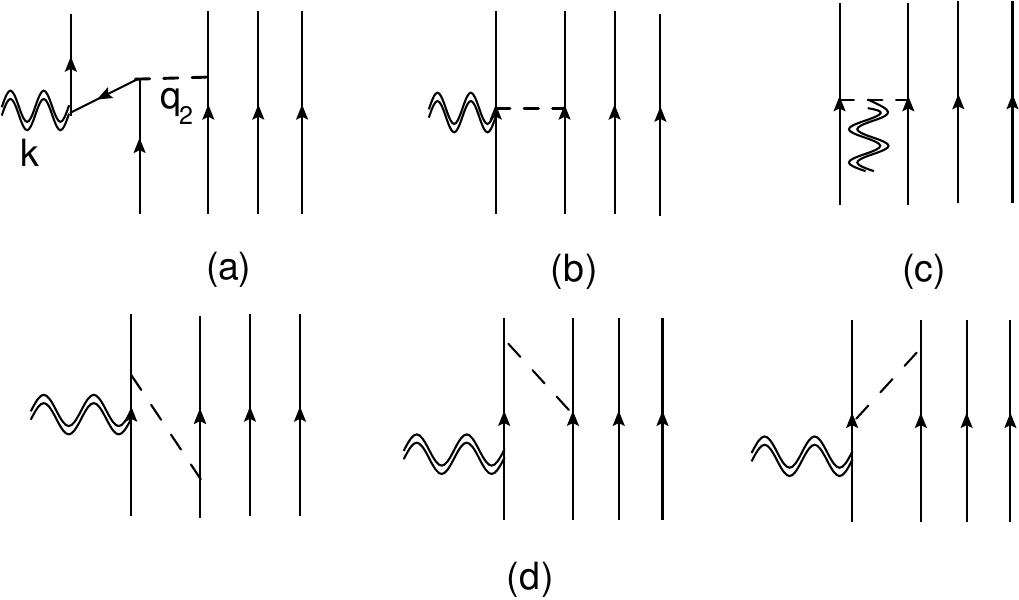}
\caption{
Exchange contributions to the EMT: (a)  The pair contribution, (b) seagull contribution, (c) pion exchange contribution, (d)  recoil plus  wave function renormalization. The solid, dashed and wavy lines represent the nucleon, pion and graviton, respectively. Note that there is no  seagull contribution for a  pseudoscalar pion-nucleon coupling.} 
\label{fig:NR_diagrams}
\end{center}
\end{figure*}

\subsection{Matrix element: $v14$ potential}
The generic form of the exchange corrections, using the Argonne potential with the variational  wave function~\eqref{eq:Jastrawsim1} are of the form
\bea
\label{eq:Tmunuv14}
\left<+\frac{k}{2} m'\bigg|T_{X\pi}^{\mu\nu}\bigg|-\frac{k}{2} m\right>
&=&\frac{1}{Z}\int d^3\vec{r}_{12}\int d^3\vec{r}_{13}\int d^3\vec{r}_{14}\frac{d^3q_2}{(2\pi)^3}
\nonumber\\
&\times&\textbf{exp}\Bigg[i\vec{q}_2.\vec{r}_{12}-i\vec{k}.(\vec{r}_1-\vec{R}_C)\Bigg]\langle\Psi_v|T_{X}^{\mu\nu}(\sigma_1,\sigma_2)\Psi_v\rangle+{\rm permutation},
\eea
where $\vec{r}_{ij}=\vec{r}_i-\vec{r}_j$. Here  $Z$ enforces the wavefunction normalization
\bea
Z=\int d^3\vec{r}_{12}\int d^3\vec{r}_{13}\int d^3\vec{r}_{14}\langle\Psi_v|\Psi_v\rangle.
\eea
\end{widetext}
Note that we have switched back to initial coordinates
 $\vec{r}_{12}$, $\vec{r}_{13}$ and $\vec{r}_{14}$ as integration  variables, which
 tie to the Jacobi coordinates (\ref{eq:Jacocorr}) by translational invariance, modulo a constant Jacobian. The latter can be eliminated in the ratio (\ref{eq:Tmunuv14}). To evaluate
 the multi-dimensional integral in Eq.~(\ref{eq:Tmunuv14}), 
 we will use the Monte Carlo method described in~\cite{Lomnitz-Adler:1981dmh, Wiringa:1991kp}.

\begin{figure*}[htbp] 
\begin{minipage}[b]{.45\linewidth}
\hspace*{-0.5cm}\includegraphics[width=1.12\textwidth, height=5.5cm]{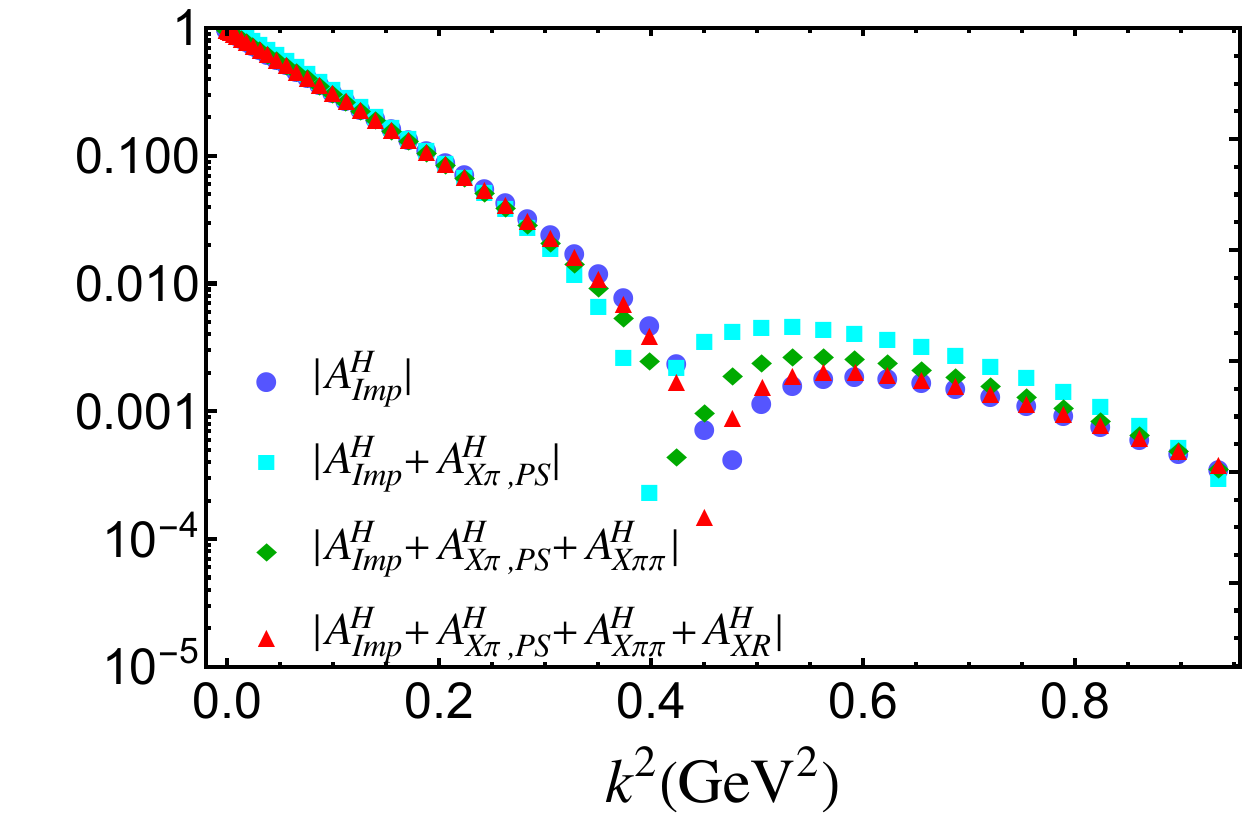}   
 \vspace{-10pt}
\end{minipage}
\hfill
\begin{minipage}[t]{.45\linewidth}   
\hspace*{-0.85cm} \includegraphics[width=1.1\textwidth, height=5.5cm]{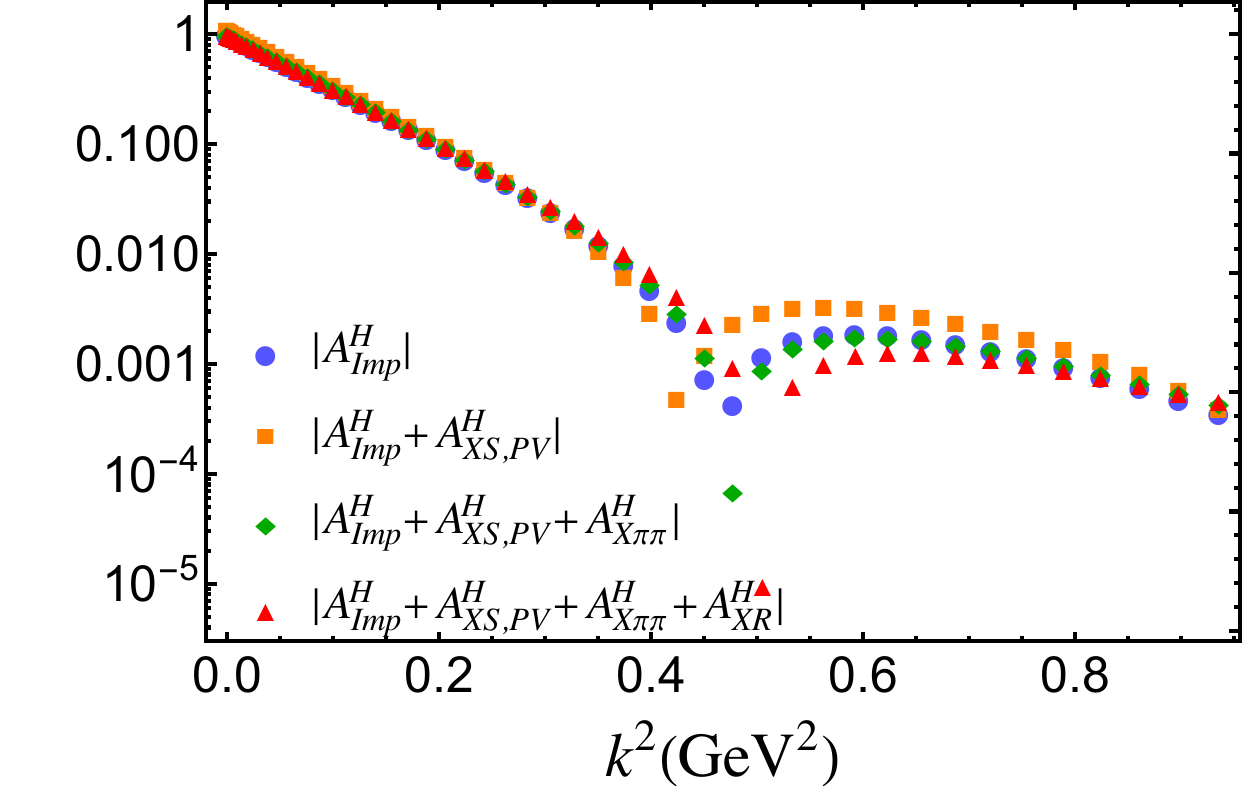}  
   \vspace{0pt}
\end{minipage} 
\begin{minipage}[t]{.45\linewidth}
\hspace*{-0.5cm}\includegraphics[width=1.12\textwidth, height=5.5cm]{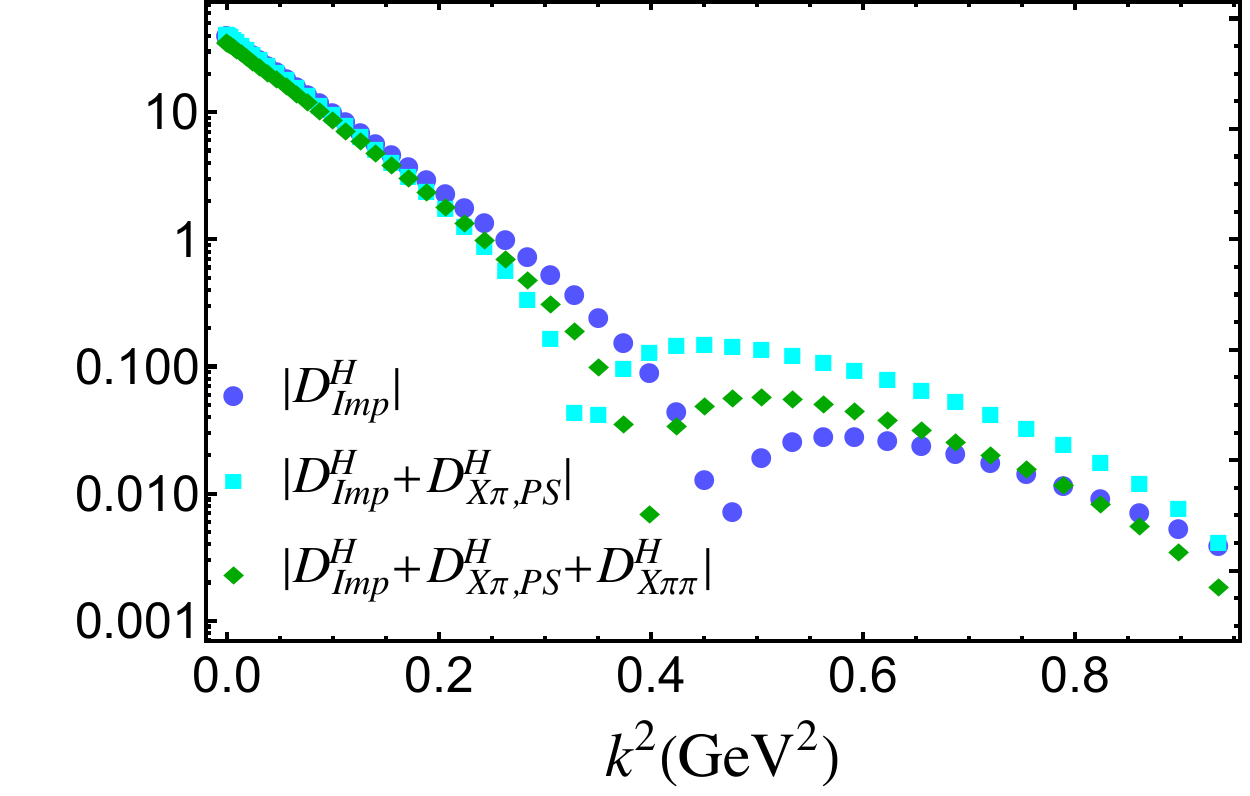}   
 \vspace{0pt}
\end{minipage}
\hfill
\begin{minipage}[t]{.45\linewidth}   
\hspace*{-0.85cm} \includegraphics[width=1.1\textwidth, height=5.5cm]{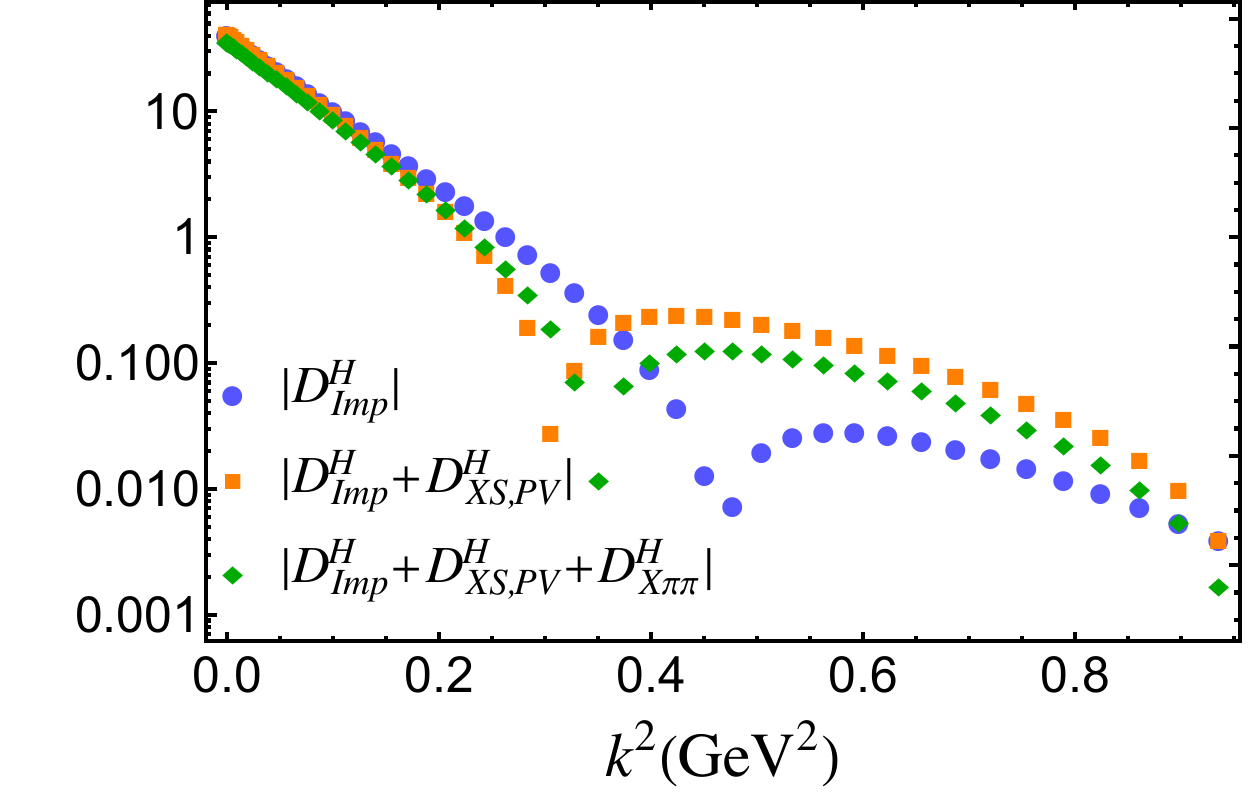}  
   \vspace{-12pt}
\end{minipage}
\caption{Helium-4 GFFs with PS coupling (left) and PV coupling (right) using the K-Harmonics method: ``imp" refers to the impulse approximation,   ``$X\pi$", ``$XS$", ``$X\pi\pi$" and ``XR" refer to the exchange contributions in~(Fig.~\ref{fig:NR_diagrams}a),  seagull contribution in~(Fig.~\ref{fig:NR_diagrams}b),  pion exchange contribution in~(Fig.~\ref{fig:NR_diagrams}c), and recoil contribution in~(Fig.~\ref{fig:NR_diagrams}d), respectively.  }
\label{fig:ex_resultKH}
\end{figure*}

\section{Matrix element of EMT in Helium-4}
\label{SECIV}
To extract the non-relativistic contributions to the exchange currents, we will make use of the diagrams in~Fig.~\ref{fig:NR_diagrams}, with the positive nucleon contribution in (a)
subtracted, and the recoil contribution and wavefunction renormalization
in (d) retained. The wiggly line refers to the insertion of the EMT,
and the dashed line to the pion exchange contribution
using both pseudo-vector coupling and pseudo-scalar coupling. 
Again, the explicit operator form for the EMT insertion $T_X$ for nucleon pairs in the context of the impulse approximation,
is given in~\cite{He:2023ogg} and to which we refer for completeness.
Instead, we will give the explicit integral expressions for the  form factors for each contribution, using  the K-Harmonic Method only. Those from the 
Argonne $v14$ will be evaluated numerically using Eq.~(\ref{eq:Tmunuv14}) by   Monte Carlo sampling, with only the final results given  in section~\ref{SECV}.

\subsection{Pair contribution}
The EMT operator for the pair diagrams in~Fig.~\ref{fig:NR_diagrams}a with pseudo-scalar coupling and pseudo-vector coupling,  in the leading non-relativistic reduction read~\cite{He:2024vzz}
\begin{widetext}
\begin{subequations}
\label{eq:Tmunupair}
\bea
\label{eq:Tmunupair_PS00}
T_{X\pi,PS}^{00}(k)&=&(A(k)+B(k))\frac{g_{\pi N}^2}{(2m_N)^2}\tau_1\cdot\tau_2\frac{\vec{\sigma}_1.\vec{k}\vec{\sigma}_2.\vec{q_2}}{w_{q_2}^2}+(1\leftrightarrow 2)+{\cal O}\bigg(\frac{g_{\pi N}^2}{m_N^4}\bigg)\\
\label{eq:Tmunupair_PSij}
T_{X\pi,PS}^{ij}(k)&=&-A(k)\tau_1\cdot\tau_2\frac{g_{\pi N}^2}{(2m_N)^2}\frac{q_2^i \sigma_1^j+q_2^j \sigma_1^i}{2 w_{q_2}^2}\vec{\sigma}_2.\vec{q}_2+(1\leftrightarrow 2)+{\cal O}\bigg(\frac{g_{\pi N}^2}{m_N^4}\bigg), \\
\label{eq:Tmunupair_PV}
T_{X\pi,PV}^{00, ij}(k)&=&{\cal O}\bigg(\frac{g_{\pi N}^2}{m_N^4}\bigg)
\\ 
\eea
\end{subequations}
\end{widetext}
with the pion energy $w_{q_2}=\sqrt{q_2^2+m_\pi^2}$.
The GFFs $A$ and $D$ for Helium-4 are  related to the matrix elements of $T^{00}$ and $T^{ij}$ through
\bea
\left<+\frac{k}{2} \bigg|T_{X\pi}^{00}\bigg|-\frac{k}{2} \right>&=&(m_\alpha+\frac{k^2}{4m_\alpha})A^H_{X\pi}+\frac{k^2}{4m_\alpha}D^H_{X\pi},
\nonumber\\
\left<+\frac{k}{2} \bigg|T_{X\pi}^{ij}\bigg|-\frac{k}{2} \right>&=&\frac{k_ik_j-\delta^{ij}k^2}{4m_\alpha}D^{H}_{X\pi} + \delta^{ij}D^{H}_{1,X\pi}.
\nonumber\\
\eea
Note that the term including $D^{H}_{1,X\pi}$ breaks down the conservation law $k^i T_{X\pi}^{ij}=0$. As noted in~\cite{He:2024vzz}, all contributions in Fig~.\ref{fig:NR_diagrams} need to be summed up, to uphold the conservation law. In particular, the contributions $\delta^{ij}$ will be cancelled out. Therefore,  we will not keep such a contribution in our following calculation. With this in mind, the corresponding GFFs for the pair diagrams with PS and PV coupling can be expressed as 
\begin{widetext}
\begin{subequations}
\bea
\label{eq:FF_AXpairPS}
A^H_{X\pi,PS}&=&-\frac{\vec{k}^2}{\vec{k}^2+4m_\alpha^2}D_{X\pi,PS}^H + 6\frac{(4\pi)^2g_{\pi N}^2}{(2m_N)^2}m_\pi^2\int_0^\infty dR \int_0^\frac{\pi}{2} d\theta \int_0^\frac{\pi}{2} d\phi \,{\mathbb  V}\, j_1\left(\frac{k R cos\theta}{\sqrt{2}}\right) \frac{k u^2 Y_1^H}{6m_N}
\nonumber\\ &\times&(A(k)+B(k))\langle\Phi|\tau_1\cdot\tau_2\,\sigma_1\cdot\sigma_2|\Phi\rangle,
\\
\label{eq:FF_DXpairPS}
D^H_{X\pi,PS}&=&6\frac{(4\pi)^2g_{\pi N}^2}{(2m_N)^2}m_\pi^2 \int_0^\infty dR \int_0^\frac{\pi}{2} d\theta \int_0^\frac{\pi}{2} d\phi \,{\mathbb  V}\, j_2\left(\frac{k R cos\theta}{\sqrt{2}}\right) \frac{-8m_\alpha m_\pi u^2 Y_2^H}{3 k^2 }A(k)
\nonumber\\ &\times&\langle\Phi|\tau_1\cdot\tau_2\,\sigma_1\cdot\sigma_2|\Phi\rangle,
\\
\label{eq:FF_ADXpairPV}
A^H_{X\pi,PV}&=&D^H_{XS,PV}={\cal O}\bigg(\frac{g_{\pi N}^2}{m_N^4}\bigg),
\eea
\end{subequations}
where $u(R)$  is the reduced hyper-radius wavefunction in Eq.~(\ref{eq:sch_Hel_new}), and factor ${\mathbb V}$ is defined as
\bea
{\mathbb V}&=&sin\theta^5 cos\theta^2sin\phi^2 cos\phi^2 j_0\left(\frac{kR\,sin\theta cos\phi}{\sqrt{6}}\right)j_0\left(\frac{kR\,sin\theta sin\phi}{2\sqrt{3}}\right),
\eea
with the  shorthand Yukawa potentials,
\bea
Y_0^H&=&e^{-m_\pi \sqrt{2}R\, cos\theta}/(m_\pi \sqrt{2}R\, cos\theta),\nonumber\\
\bar{Y}_1^H&=&(1+1/(m_\pi \sqrt{2}R\,cos\theta))Y_0^H,\nonumber\\
Y_2^H&=&(3/(m_\pi \sqrt{2}R\, cos\theta)^2+3/(m_\pi \sqrt{2}R\, cos\theta)+1)Y_0^H.
\eea
\end{widetext}
In the $0^{++}$ Helium-4 ground state, we can use Eq.~(\ref{eq:spispwf}) to obtain
\bea
&&\langle\Phi|\tau_1\cdot\tau_2\, \sigma_1\cdot\sigma_2|\Phi\rangle=-\frac{315}{32\pi^4}.
\eea

\subsection{Seagull contribution}
For the Seagull in~Fig~.\ref{fig:NR_diagrams}(b), the EMT operator can be written as~\cite{He:2024vzz}
\begin{subequations}
\label{eq:TmunuS}
\bea
\label{eq:TmunuS_PV00}
T_{XS,PV}^{00}&=&{\cal O}\bigg(\frac{g_{\pi N}^2}{m_N^4}\bigg),
\\
\label{eq:TmunuS_PVij}
T_{XS,PV}^{ij}&=&-\tau_1\cdot\tau_2\frac{g_{\pi N}^2}{(2m_N)^2}\frac{q_2^i \sigma_1^j+q_2^j \sigma_1^i}{2 w_{q_2}^2}\vec{\sigma}_2.\vec{q}_2\nonumber\\
&&\qquad+(1\leftrightarrow 2)
+{\cal O}\bigg(\frac{g_{\pi N}^2}{m_N^4}\bigg).
\eea
\end{subequations}
There is no contribution from the seagull diagram for the PS coupling case. Also the contribution to  $T^{ij}_{XS}$ from  the PV coupling is equivalent to the contribution to $T^{ij}_{X\pi}$ from the pair diagram for the PS coupling,with the exception of the extra nucleon GFF A(k) in Eq.~(\ref{eq:Tmunupair_PSij}).
The GFFs can be obtained using the following relations
\bea
\left<+\frac{k}{2} \bigg|T_{XS}^{00}\bigg|-\frac{k}{2} \right>&=&(m_\alpha+\frac{k^2}{4m_\alpha})A^H_{XS}+\frac{k^2}{4m_\alpha}D^H_{XS}
\nonumber\\
&=&{\cal O}\bigg(\frac{g_{\pi N}^2}{m_N^4}\bigg),
\nonumber\\
\left<+\frac{k}{2} \bigg|T_{XS}^{ij}\bigg|-\frac{k}{2} \right>&=&\frac{k_ik_j-\delta^{ij}k^2}{4m_\alpha}D^{H}_{XS},
\eea
where 
\begin{widetext}
\begin{subequations}
\bea
\label{eq:FF_ADXSPS}
A^H_{XS,PS}&=&D^H_{XS,PS}=0,
\\
\label{eq:FF_AXSPV}
A^H_{XS,PV}&=&-\frac{\vec{k}^2}{\vec{k}^2+4m_\alpha^2}D_{XS,PV}^H,
\\
\label{eq:FF_DXSPV}
D^H_{XS,PV}&=&6\frac{(4\pi)^2g_{\pi N}^2}{(2m_N)^2}m_\pi^2 \int_0^\infty dR \int_0^\frac{\pi}{2} d\theta \int_0^\frac{\pi}{2} d\phi {\mathbb V}\, j_2\left(\frac{k R cos\theta}{\sqrt{2}}\right) \frac{-8m_\alpha m_\pi u^2 Y_2^H}{3 k^2 }
\nonumber\\ &\times&\langle\Phi|\tau_1\cdot\tau_2\,\sigma_1\cdot\sigma_2|\Phi\rangle.
\eea
\end{subequations}

\begin{figure*}[htbp] 
\begin{minipage}[b]{.45\linewidth}
\hspace*{-0.3cm}\includegraphics[width=1.1\textwidth, height=5.5cm]{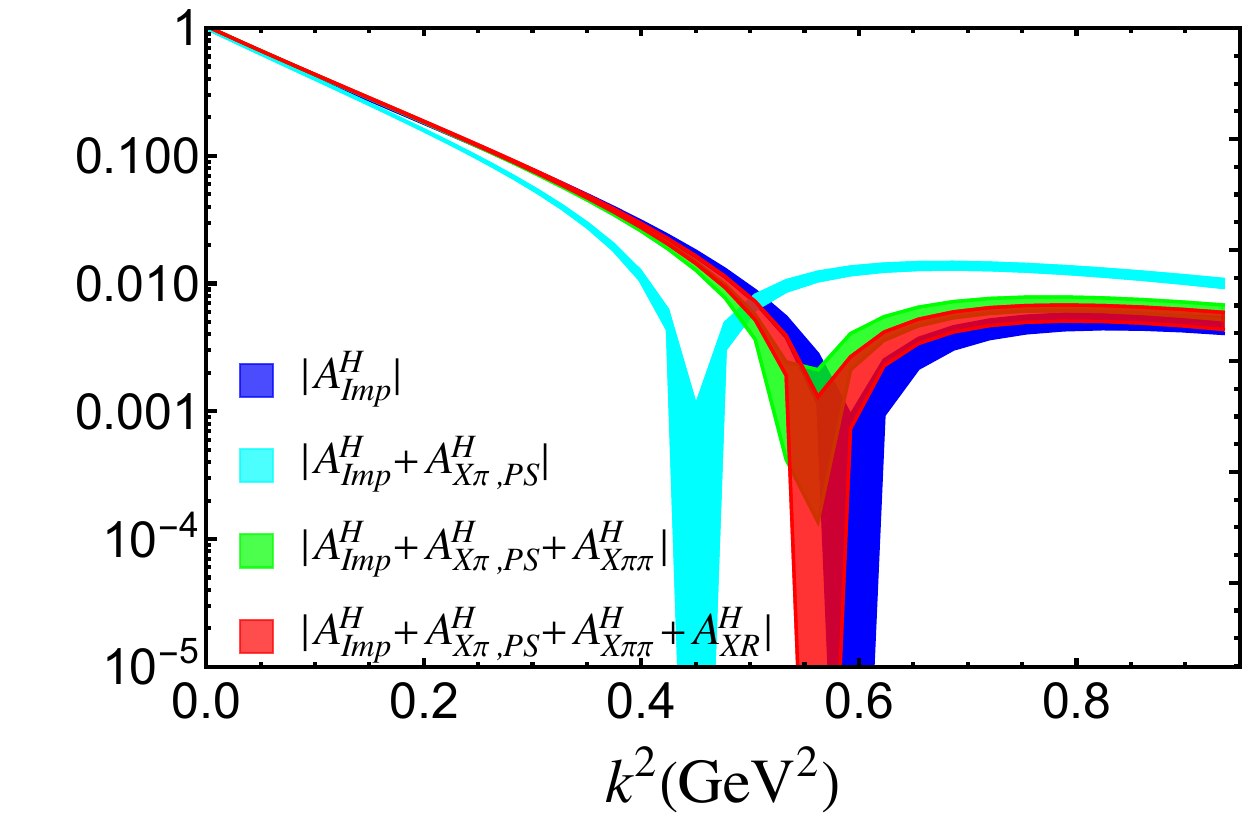}   
 \vspace{0pt}
\end{minipage}
\hfill
\begin{minipage}[b]{.45\linewidth}   
\hspace*{-0.85cm} \includegraphics[width=1.1\textwidth, height=5.5cm]{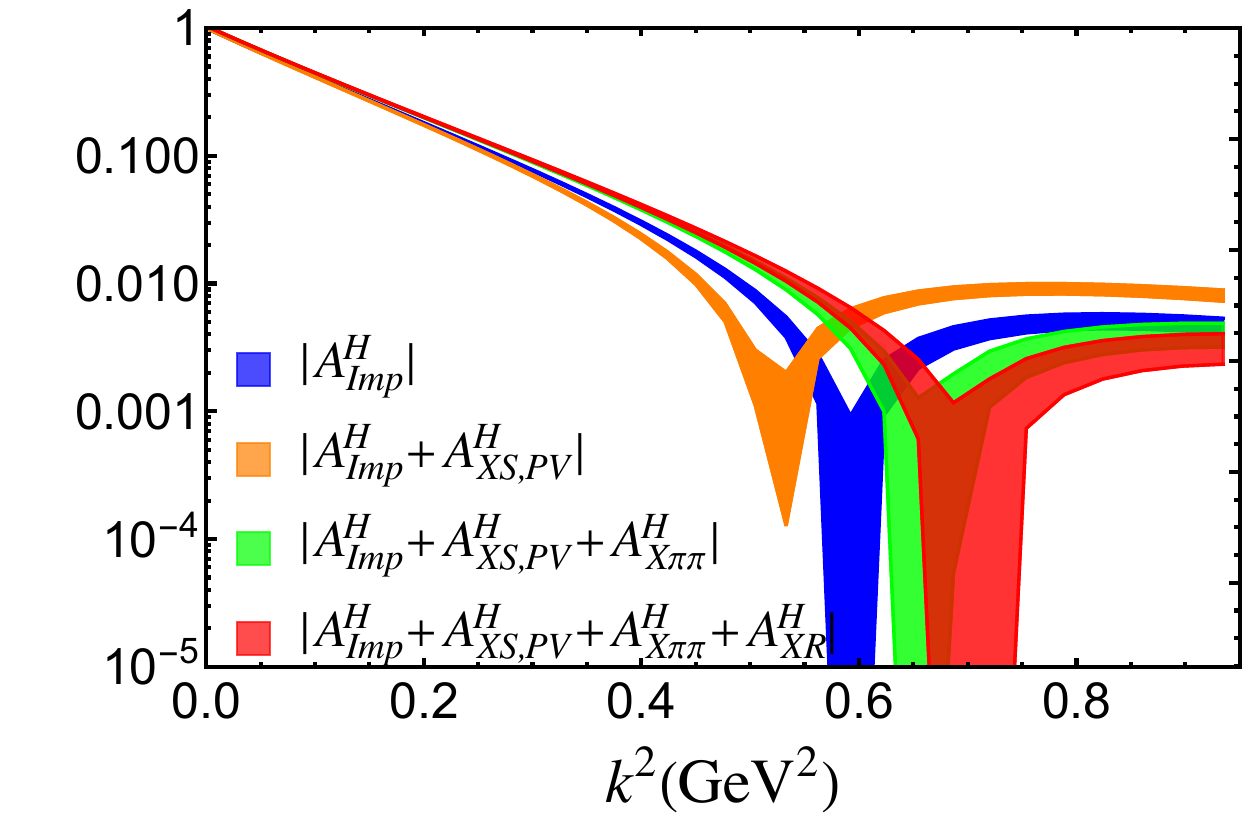} 
   \vspace{0pt}
\end{minipage}  
\\[-0.4cm]
\begin{minipage}[t]{.45\linewidth}
\hspace*{-0.5cm}\includegraphics[width=1.12\textwidth, height=5.5cm]{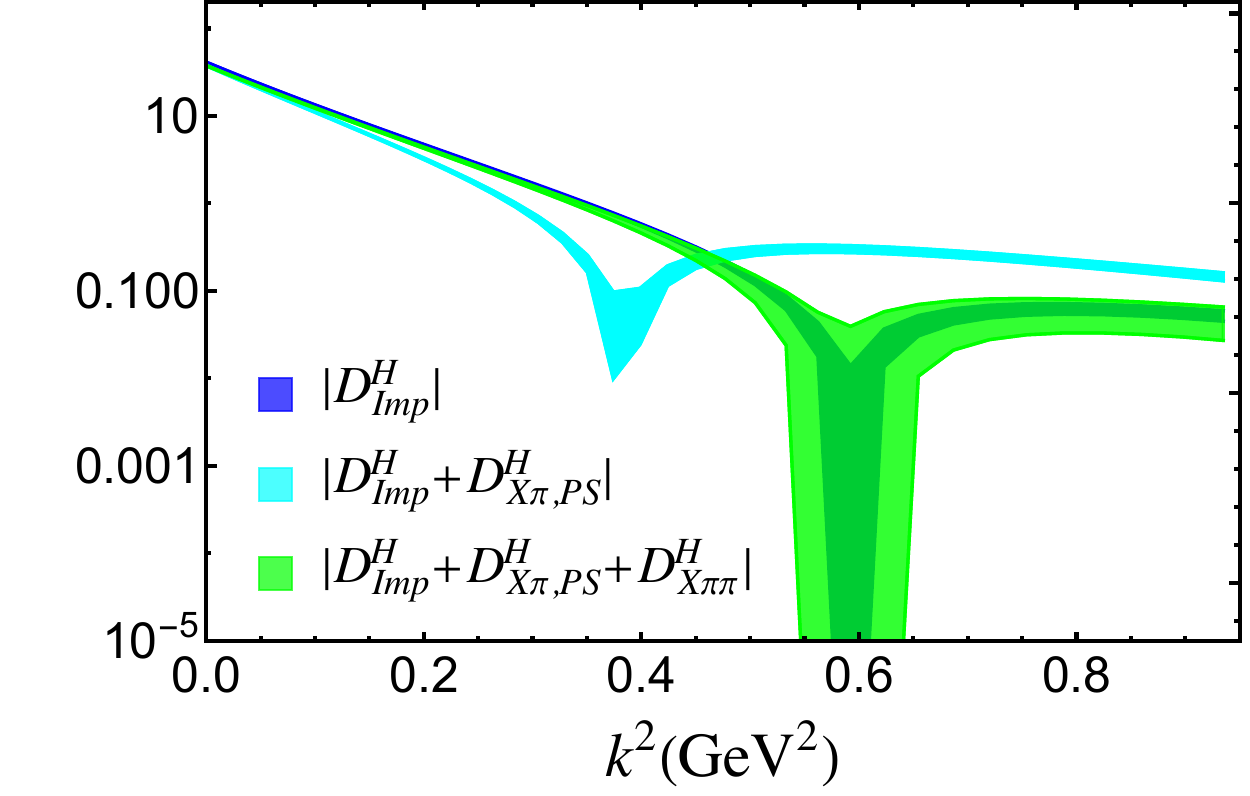}   
 \vspace{0pt}
\end{minipage}
\hfill
\begin{minipage}[t]{.45\linewidth}   
\hspace*{-0.85cm} \includegraphics[width=1.1\textwidth, height=5.5cm]{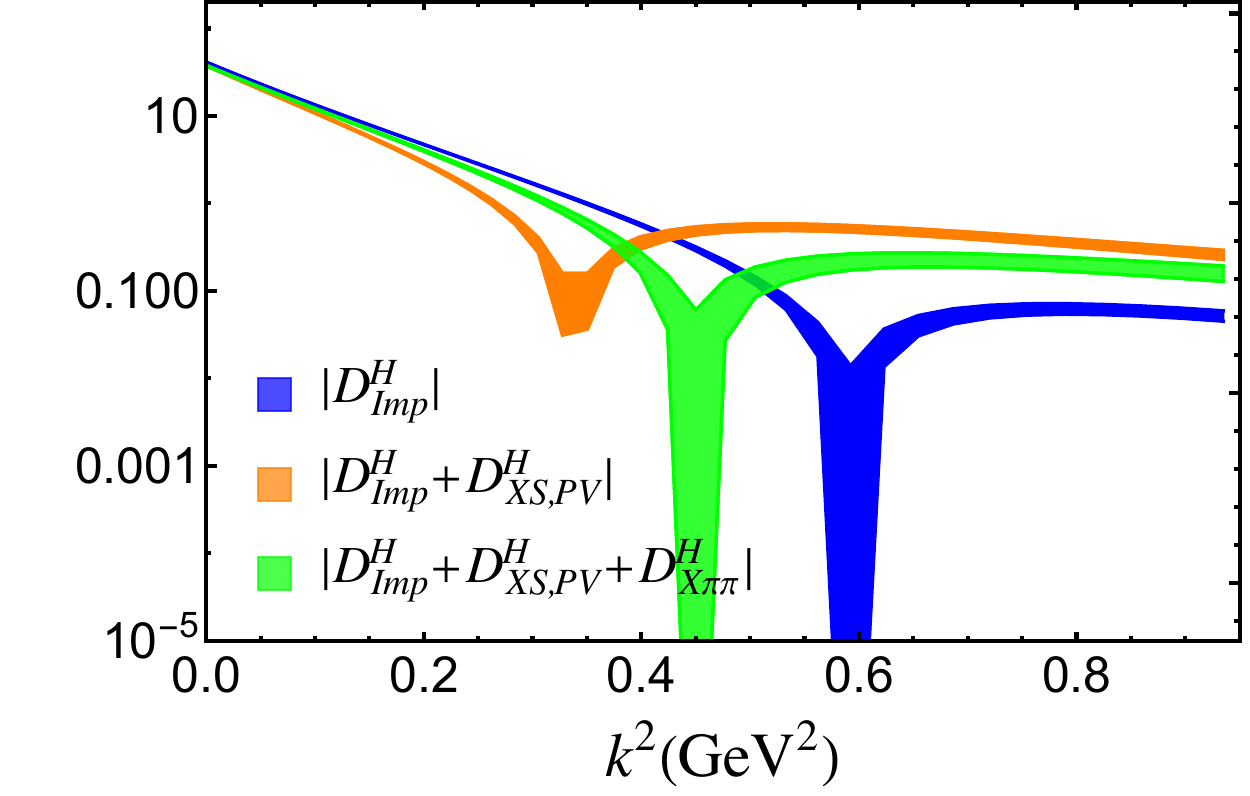}  
   \vspace{0pt}
\end{minipage} 
\caption{ Helium-4 GFFs with PS coupling (left panel) and PV coupling (right panel),  using the Argonne $v14$ potential via a Monte Carlo method. The bands represent the statistical uncertainty.}
\label{fig:ARGONNE}
\end{figure*}

\subsection{Pion exchange contribution}
The EMT exchange contributions stemming from the one-pion exchange in the non-relativistic reduction, are equivalent whether we use  the PS or PV couplings. More specifically, the pertinent operators are~\cite{He:2024vzz}
\begin{subequations}    
\label{eq:TmunuPi}
\bea
\label{eq:TmunuPi_00}
T_{X\pi\pi}^{00}&=&-\frac{g_{\pi N}^2}{(2m_N)^2}\tau_1\cdot\tau_2\frac{\vec{k}^2 \vec{\sigma_1}.\vec{q_1} \vec{\sigma_2}.\vec{q_2}}{2 w_{q_1}^2 w_{q_2}^2}T_{1\pi}(k)+\frac{g_{\pi N}^2}{(2m_N)^2}\tau_1\cdot\tau_2\frac{\sigma_1.q_1 \sigma_2.q_2 }{2}\left(\frac{1}{w_{q_2}^2}+\frac{1}{w_{q_1}^2}\right)
\nonumber\\
&+&{\cal O}\bigg(\frac{g_{\pi N}^2}{m_N^4}\bigg),
\\
\label{eq:TmunuPi_ij}
T_{X\pi\pi}^{ij}&=&\frac{g_{\pi N}^2}{(2m_N)^2}\tau_1\cdot\tau_2\left(\frac{(\vec{k}^2\delta^{ij}-k^ik^j) \vec{\sigma_1}.\vec{q_1} \vec{\sigma_2}.\vec{q_2}}{2 w_{q_1}^2 w_{q_2}^2}T_{1\pi}(k)+\frac{(q_1^i+q_2^i)(q_1^j+q_2^j)\vec{\sigma_1}.\vec{q_1} \vec{\sigma_2}.\vec{q_2}}{2 w_{q_1}^2 w_{q_2}^2}T_{2\pi}(k)\right)
\nonumber\\
&-&\delta^{ij}\frac{g_{\pi N}^2}{(2m_N)^2}\tau_1\cdot\tau_2\frac{\sigma_1.q_1 \sigma_2.q_2  \left(\vec{q}_1^2+\vec{q}_2^2+2m_\pi^2\right)}{2 w_{q_1}^2 w_{q_2}^2}+{\cal O}\bigg(\frac{g_{\pi N}^2}{m_N^4}\bigg),
\eea
\end{subequations}
with the identical matrix elements
\bea
\left<+\frac{k}{2} \bigg|T_{X\pi\pi}^{00}\bigg|-\frac{k}{2} \right>&=&(m_\alpha+\frac{k^2}{4m_\alpha})A^H_{X\pi\pi}+\frac{k^2}{4m_\alpha}D^H_{X\pi\pi},
\nonumber\\
\left<+\frac{k}{2} \bigg|T_{X\pi\pi}^{ij}\bigg|-\frac{k}{2} \right>&=&\frac{k_ik_j-\delta^{ij}k^2}{4m_\alpha}D^{H}_{X\pi\pi},
\eea
where 
\begin{subequations}
\bea
\label{eq:FF_A_pi}
A_{X\pi\pi}^H&=&-\frac{\vec{k}^2}{\vec{k}^2+4m_\alpha^2}D_{X\pi\pi}^H+6\frac{(4\pi)^2g_{\pi N}^2}{(2m_N)^2}T_{1\pi}(k)\int_{-\frac{1}{\sqrt{2}}}^{\frac{1}{\sqrt{2}}} dt\int_0^\infty dR \int_0^\frac{\pi}{2} d\theta \int_0^\frac{\pi}{2} d\phi {\mathbb V}\Bigg\{
\nonumber\\
&&\frac{k L_\pi u^2  \left(\sqrt{2} k^2 R t^2 Y_0^{H,\pi\pi} cos\theta+3 L_\pi \bar{Y}_1^{H,\pi\pi}\right)}{12 t\, m_{\alpha }}j_1(k t R\,  cos\theta )-\frac{k^2 L_\pi^2 R  u^2 \bar{Y}_1^{H,\pi\pi} cos\theta j_2(k t R\,  cos\theta )}{12 m_{\alpha }}
\nonumber\\
&-&\frac{k^2 u^2 Y_0^{H,\pi\pi} \left(k^2 R  \left(2 t^2-1\right) cos\theta+6 \sqrt{2} L_\pi\right) j_0(k t R  cos\theta)}{48 m_{\alpha }}
\Bigg\}\langle\Phi|\tau_1\cdot\tau_2\,\sigma_1\cdot\sigma_2|\Phi\rangle
\nonumber\\
&+&6\frac{(4\pi)^2g_{\pi N}^2}{(2m_N)^2}\int dR \int_0^\frac{\pi}{2} d\theta \int_0^\frac{\pi}{2} d\phi V\frac{m_\pi^2 u^2  }{12  k m_\alpha R}\Bigg\{-2j_1\left(\frac{k R cos\theta}{\sqrt{2}}\right) \left(k^2 r \bar{Y}^H_1+3\sqrt{2} m_\pi Y^H_2 sec\theta\right)
\nonumber\\
&+&2k m_\pi R Y^H_2 j_2\left(\frac{k R cos\theta}{\sqrt{2}}\right)+3\sqrt{2} k \bar{Y}^H_1 sec\theta j_0\left(\frac{k r cos\theta}{\sqrt{2}}\right)\Bigg\}\langle\Phi|\tau_1\cdot\tau_2\,\sigma_1\cdot\sigma_2|\Phi\rangle,
\\
\label{eq:FF_D_pi}
D_{X\pi\pi}^H&=&6\frac{(4\pi)^2g_{\pi N}^2}{(2m_N)^2}T_{1\pi}(k)\int_{-\frac{1}{\sqrt{2}}}^{\frac{1}{\sqrt{2}}} dt\int_0^\infty dR \int_0^\frac{\pi}{2} d\theta \int_0^\frac{\pi}{2} d\phi\, {\mathbb V}\nonumber\\
&\times&\Bigg\{-\frac{m_\alpha u^2 Y_0^{H,\pi\pi}\left[k^2 R \left(2 t^2-1\right)cos\theta+6\sqrt{2} L_\pi\right] j_0(k t R\,cos\theta)}{12 }
\nonumber\\
&+&\frac{m_\alpha L_\pi u^2 j_1(k t R\,cos\theta) \left(\sqrt{2} k^2 R t^2 Y_0^{H,\pi\pi}cos\theta+3 L_\pi \bar{Y}_1^{H,\pi\pi}\right)}{3  k t}
\nonumber\\
&-&\frac{m_\alpha L_\pi^2 R u^2 \bar{Y}_1^{H,\pi\pi}cos\theta  j_2(k t R\,cos\theta)}{3 }\Bigg\}\langle\Phi|\tau_1\cdot\tau_2\,\sigma_1\cdot\sigma_2|\Phi\rangle
\nonumber\\
&+&6\frac{(4\pi)^2g_{\pi N}^2}{(2m_N)^2}T_{2\pi}(k)\int_{-\frac{1}{\sqrt{2}}}^{\frac{1}{\sqrt{2}}} dt\int_0^\infty dR \int_0^\frac{\pi}{2} d\theta \int_0^\frac{\pi}{2} d\phi {\mathbb V}\Bigg\{
\nonumber\\
&-&\frac{u^2 m_{\alpha }j_0(k t R cos\theta)}{54 k^2} [R  Y_0^{H,\pi\pi}cos\theta \left(9 k^4 t^2 \left(1-2 t^2\right)+40 L_\pi^4\right)\nonumber\\
&+&2 \sqrt{2} L_\pi \left(10 L_\pi^2 (Y_0^{H,\pi\pi}+5 Y_2^{H,\pi\pi})-63 k^2 t^2 Y_0^{H,\pi\pi}\right)]
\nonumber\\
&+&\frac{L_\pi t u^2 m_{\alpha } j_1(k t R cos\theta)}{15 k} [-\sqrt{2} R  cos\theta \left(5 k^2 \left(4 t^2-1\right) Y_0^{H,\pi\pi}-28 L_\pi^2 Y_2^{H,\pi\pi}\right)-210 L_\pi \bar{Y}_1^{H,\pi\pi}]
\nonumber\\
&-&\frac{L_\pi^2 u^2 \sec^2\theta m_{\alpha }j_2(k t R  cos\theta) }{756 k^4 R^2 t^2}\Big[452 k^2 L_\pi^2 R^3 t^2 Y_0^{H,\pi\pi} cos(3 \theta)  
+3 R  cos\theta \Big(4 L_\pi^2 Y_0^{H,\pi\pi} \left(113 k^2 R ^2 t^2-700\right)
\nonumber\\
&+&63 k^4 R^2 t^2 \left(1-12 t^2\right) \bar{Y}_1^{H,\pi\pi}\Big)
+452 \sqrt{2} k^2 L_\pi R^2 t^2 Y_0^{H,\pi\pi}+4 \sqrt{2} k^2 L_\pi R^2 t^2 cos(2 \theta) (113 Y_0^{H,\pi\pi}+124 Y_2^{H,\pi\pi})
\nonumber\\
&+&496 \sqrt{2} k^2 L_\pi R^2 t^2 Y_2^{H,\pi\pi}+63 k^4 R^3 t^2 \bar{Y}_1^{H,\pi\pi}cos (3 \theta) -756 k^4 R^3 t^4 \bar{Y}_1^{H,\pi\pi} cos(3\theta )
\nonumber\\
&-&4200 \sqrt{2} L_\pi Y_0^{H,\pi\pi}-21000 \sqrt{2} L_\pi Y_2^{H,\pi\pi}\Big]
\nonumber\\
&-&\frac{4 \sqrt{2} L_\pi^3  t u^2 Y_2^{H,\pi\pi}R\, cos\theta m_{\alpha }j_3(k t R  cos \theta)}{5 k}
\nonumber\\
&-&\frac{10 L_\pi^3 u^2 m_{\alpha } j_4(k t R\, cos\theta ) \left(2 L_\pi R  Y_0^{H,\pi\pi} cos \theta+\sqrt{2} (Y_0^{H,\pi\pi}+5 Y_2^{H,\pi\pi})\right)}{63 k^2}
\Bigg\}\langle\Phi|\tau_1\cdot\tau_2\,\sigma_1\cdot\sigma_2|\Phi\rangle.
\eea
Here $L_\pi=\left[m_\pi^2+(\frac14-\frac{t^2}{2})\vec{k}^2\right]^{1/2}$ and the shorthand Yukawa potentials are expressed as
\bea
Y_0^{H,\pi\pi}&=&e^{-L_\pi \sqrt{2}R\, cos\theta}/(L_\pi \sqrt{2}R\, cos\theta),\nonumber\\
\bar{Y}_1^{H,\pi\pi}&=&(1+1/(L_\pi \sqrt{2}R\,cos\theta))Y_0^{H,\pi\pi},\nonumber\\
Y_2^{H,\pi\pi}&=&(3/(L_\pi \sqrt{2}R\, cos\theta)^2+3/(L_\pi \sqrt{2}R\, cos\theta)+1)Y_0^{H,\pi\pi}.
\eea
\end{subequations}

\subsection{Recoil contribution}
The exchange operators stemming from the recoil terms after non-relativistic reduction are also equivalent whether we use the PS or PV coupling, they can be expressed as~\cite{He:2024vzz}
\begin{subequations} 
\label{eq:TmunuR}
\bea
\label{eq:TmunuR00}
T^{00}_{XR}
&=&-\frac{g_{\pi N}^2}{(2m_N)^2}\tau_1\cdot\tau_2A(k)\frac{q_2.k \sigma_1.q_2 \sigma_2.q_2}{2 w_{q_2}^4}+(1\leftrightarrow 2)+{\cal O}\bigg(\frac {g_{\pi N}^2}{m_N^4}\bigg),
\\
\label{eq:TmunuRij}
T^{ij}_{XR}&=&\bigg(\frac {g_{\pi N}^2}{m_N^4}\bigg).
\eea
\end{subequations}
with the corresponding matrix elements
\bea
\left<+\frac{k}{2} \bigg|T_{XR}^{00}\bigg|-\frac{k}{2} \right>&=&(m_\alpha+\frac{k^2}{4m_\alpha})A^H_{XR}+\frac{k^2}{4m_\alpha}D^H_{XR},
\nonumber\\
\left<+\frac{k}{2} \bigg|T_{X}^{ij}\bigg|-\frac{k}{2} \right>&=&\frac{k_ik_j-\delta^{ij}k^2}{4m_\alpha}D^{H}_{XR}={\cal O}\bigg(\frac{g_{\pi N}^2}{m_N^4}\bigg),
\eea
and
\bea
\label{eq:FF_A_XR}
A^H_{XR}=&&6\frac{(4\pi)^2g_{\pi N}^2}{(2m_N)^2}A(k)\nonumber\\
&&\times\int_0^\infty dR \int_0^\frac{\pi}{2} d\theta \int_0^\frac{\pi}{2} d\phi {\mathbb V} \frac{k m_\pi^2 u^2 j_1\left(\frac{k R cos\theta}{\sqrt{2}}\right) (\sqrt{2}m_\pi R Y_2^Hcos\theta-5 \bar{Y}_1^H)}{ 6m_\alpha}
\langle\Phi|\tau_1\cdot\tau_2\,\sigma_1\cdot\sigma_2|\Phi\rangle.
\label{eq:FF_D_XR}
\eea

\end{widetext}

\section{Numerical results}
\label{SECV}
To summarise, the EMT exchange contributions to  Helium-4 for the PS and PV 
pion nucleon couplings are given by
\bea
\label{SUMDX}
T^{00}_{DX,PS}&=&T_{X\pi}^{00}+T_{X\pi\pi}^{00}+T^{00}_{XR},\nonumber\\
T^{ij}_{DX,PS}&=&T_{X\pi}^{00}+T_{X\pi\pi}^{00},\nonumber\\
T^{00}_{DX,PV}&=&T_{X\pi\pi}^{00}+T^{00}_{XR},\nonumber\\
T^{ij}_{DX,PV}&=&T_{XS}^{ij}+T^{ij}_{X\pi\pi}.
\eea
The main difference between the PS  and PV couplings, is the additional  contribution from the pair diagram to the $T^{00}$ component in the PS coupling case. A similar addition is noted  in the charge form factor as in~(\ref{eq:J0pair}). We also note that  
the pair diagram contribution following from the PS coupling to $T^{ij}$ , is very similar to that of the seagull diagram (absent for the PS coupling) with PV coupling. The only difference is the extra nucleon form factor $A(k)$ included in~(\ref{eq:Tmunupair_PSij}) in comparison  to~(\ref{eq:TmunuS_PVij}).

The corresponding GFFs from the exchange currents are
\bea\label{eq:FF_sum}
A^H_{X,PS}&=&A^H_{X\pi,PS}(\ref{eq:FF_AXpairPS})+A^H_{X\pi\pi}(\ref{eq:FF_A_pi})
\nonumber\\
&+&A^H_{XR}(\ref{eq:FF_A_XR})
\nonumber\\
D^H_{X,PS}&=&D^H_{X\pi,PS}(\ref{eq:FF_DXpairPS})+D^H_{X\pi\pi}(\ref{eq:FF_D_pi})\nonumber\\
A^H_{X,PV}&=&A^H_{XS,PV}(\ref{eq:FF_AXSPV})+A^H_{X\pi\pi}(\ref{eq:FF_A_pi})
\nonumber\\
&+&A_{XR}(\ref{eq:FF_A_XR})\nonumber\\
D^H_{X,PV}&=&D^H_{XS,PV}(\ref{eq:FF_DXSPV})+D^H_{X\pi\pi}(\ref{eq:FF_D_pi})
\eea
To assess the  pion GFFs in the pion exchange current contribution, we  borrow their parameterization from the recent Lattice calculation in~\cite{Hackett:2023nkr},
\bea\label{eq:GFFpiLattice}
T_{1\pi}^g(k)=\frac{0.596}{1+\frac{\vec{k}^2}{0.677~\text{GeV}^2}} \nonumber\\
T_{1\pi}^q(k)=\frac{0.304}{1+\frac{\vec{k}^2}{1.44~\text{GeV}^2}} \nonumber\\
T_{2\pi}^g(k)=\frac{0.546}{1+\frac{\vec{k}^2}{1.129~\text{GeV}^2}} \nonumber\\
T_{2\pi}^q(k)=\frac{0.481}{1+\frac{\vec{k}^2}{1.262~\text{GeV}^2}} 
\eea

In Fig.~\ref{fig:ex_resultKH} we show the Helium-4 $A^H,D^H$ 
GFFs  using the  pion-nucleon PS coupling (left) and the pion-nucleon PV coupling (right) with the K-Harmonic method. The results from the impulse approximation (blue-dots) are labeled by "$Imp$" are from~\cite{He:2023ogg}. The various contributions
 labeled by  ``$X\pi$", ``$XS$", ``$X\pi\pi$", ``$XR$" 
 stem from  the exchange contributions illustrated in~Fig.~\ref{fig:NR_diagrams}a for the pair contribution, in~Fig.~\ref{fig:NR_diagrams}b for the  seagull contribution, in~Fig.~\ref{fig:NR_diagrams}c for the pi-pi contribution and in~Fig.~\ref{fig:NR_diagrams}d for the recoil contribution, respectively.  The total corrections from the exchange currents using the PS coupling (red-triangle),
do not yield qualitative changes to the GFFs $A^H$ from the impulse approximation (blue-dot). However, the contribution of the exchange currents with  PV coupling yield lowe results compared to the impulse approximation. The difference is due to the pair contribution in Fig.~\ref{fig:NR_diagrams}a,  whose contribution is higher order in the PV coupling. The contributions of the exchange currents to $D^H$ are comparable in the PS and PV case, since the only difference is the extra nucleon form factor $A(k)$ included in $D^H_{X\pi,PS}$, but absent in $D^H_{XS,PV}$. Both shift the  diffractive minima to the left of the  impulse approximation.

In Fig.~\ref{fig:ARGONNE} we show the Helium-4 $A^H,D^H$ 
GFFs  using the  pion-nucleon PS coupling (left) and the pion-nucleon PV coupling (right) with the Argonne $v14$ potential. The statistical uncertainties following from the Monte-Carlo sampling, are reflected in the colored-spreads when the various exchange contributions are added to the impulse approximation labeled by "$Imp$" (blue-spread) results. Again,  different contributions
 labeled by  ``$X\pi$", ``$XS$", ``$X\pi\pi$", ``$XR$" 
stem from  the exchange contributions for the pair contribution, the  seagull contribution, the pi-pi contribution and the recoil contribution, respectively.  Unlike the K-harmonic method, the use of the Argonne $v14$ potential with D-wave admixture, shows sensitivity to the use
of the PS (left) versus PV (right) pion-nucleon coupling. The use of the PS coupling (left) shows that the additional contributions stemming from the exchange contributions when added up (read-spread), contribute net $A,D$ GFFS  that are comparable with the impulse approximation  (blue-spread) within statistics.  
This is not the case when using the PV coupling for the Helium-4 GFFs (right), with the net result of form factor $A^H$ (red band) is lower than result with impulse approximation (blue-spread), the diffractive minima of form factor $D^H$ (green-spread) is shifted to the left in comparison to that for impulse approximation (blue-spread). Again, the difference of form factor $A^H$ stems primarily from the $T^{00}$ component of pair diagram in the PS coupling. The difference in  $D^H$ is caused by the extra nucleon GFF A(k) in~(\ref{eq:Tmunupair_PSij}) with the PS coupling, when compared to~(\ref{eq:TmunuS_PVij}) with  PV coupling. In both cases, the GFF diffractive minima from the impulse approximation, shift rightward in comparison to the K-harmonic method.

\begin{widetext}
\begin{table*}[htbp]
  \centering
\setlength{\tabcolsep}{4mm}{
  \begin{tabular}{|l|c|c|c|}
  \toprule
   r(fm) & $f_{0,1}(r)$ & $f_{1,0}(r)$ & $f_{t,0}(r)$   \\
   0.05 & 0.1644 & 0.2291 & 0   \\
    0.25 & 0.2626 & 0.3448 & 0.0339 \\
    0.50 & 0.5690 & 0.6491 & 0.0859 \\ 
    0.75 & 0.9069 & 0.9480 & 0.1264 \\ 
    1.00 & 1.0586 & 1.0735 & 0.1361 \\ 
    1.25 & 1.0512 & 1.0525 & 0.1255 \\ 
    1.50 & 0.9812 & 0.9694 & 0.1084 \\
    1.75 & 0.8965 & 0.8719 & 0.0916 \\ 
    2.00 & 0.8135 & 0.7783 & 0.0770 \\
    2.25 & 0.7369 & 0.6942 & 0.0650 \\ 
    2.50 & 0.6677 & 0.6203 & 0.0550 \\ 
    2.75 & 0.6056 & 0.5560 & 0.0469 \\ 
    3.00 & 0.5500 & 0.4999 & 0.0403 \\ 
    3.25 & 0.5002 & 0.4508 & 0.0349 \\ 
    3.50 & 0.4555 & 0.4077 & 0.0305 \\
    3.75 & 0.4153 & 0.3694 & 0.0268 \\
    4.00 & 0.3791 & 0.3354 & 0.0238 \\
    4.25 & 0.3464 & 0.3049 & 0.0212 \\
    4.50 & 0.3168 & 0.2775 & 0.0189 \\ 
    4.75 & 0.2900 & 0.2528 & 0.0169 \\
    5.00 & 0.2657 & 0.2305 & 0.0151 \\ 
    5.25 & 0.2436 & 0.2103 & 0.0136 \\
    5.50 & 0.2235 & 0.1920 & 0.0122 \\ 
    5.75 & 0.2052 & 0.1754 & 0.0109 \\
    6.00 & 0.1885 & 0.1603 & 0.0099 \\ 
    6.25 & 0.1733 & 0.1466 & 0.0089 \\
    6.50 & 0.1593 & 0.1341 & 0.0080 \\
    6.75 & 0.1466 & 0.1227 & 0.0072 \\
    7.00 & 0.1349 & 0.1123 & 0.0065 \\
    7.25 & 0.1242 & 0.1029 & 0.0059 \\
    7.50 & 0.1144 & 0.0943 & 0.0054 \\
    7.75 & 0.1054 & 0.0864 & 0.0049 \\
    8.00 & 0.0972 & 0.0792 & 0.0044 \\ 
    8.25 & 0.0896 & 0.0726 & 0.0040 \\
    8.50 & 0.0826 & 0.0666 & 0.0036 \\ 
    8.75 & 0.0762 & 0.0611 & 0.0033 \\
    9.00 & 0.0704 & 0.0561 & 0.0030 \\
    9.25 & 0.0650 & 0.0515 & 0.0027 \\
    9.50 & 0.0600 & 0.0472 & 0.0025 \\
    9.75 & 0.0554 & 0.0434 & 0.0023 \\
    10.00 & 0.0512 & 0.0388 & 0.0020 \\
  \hline
  \end{tabular}}
  \caption{The wave functions $f_{0,1}$, $f_{1,0}$, $f_{t,0}$ solutions to Eq.~(\ref{eq:seq}).}
  \label{tab:wfslov}
\end{table*}
\end{widetext}

\begin{figure*}[htbp] 
\begin{minipage}[b]{.45\linewidth}
\hspace*{-0.3cm}\includegraphics[width=1.1\textwidth, height=5.5cm]{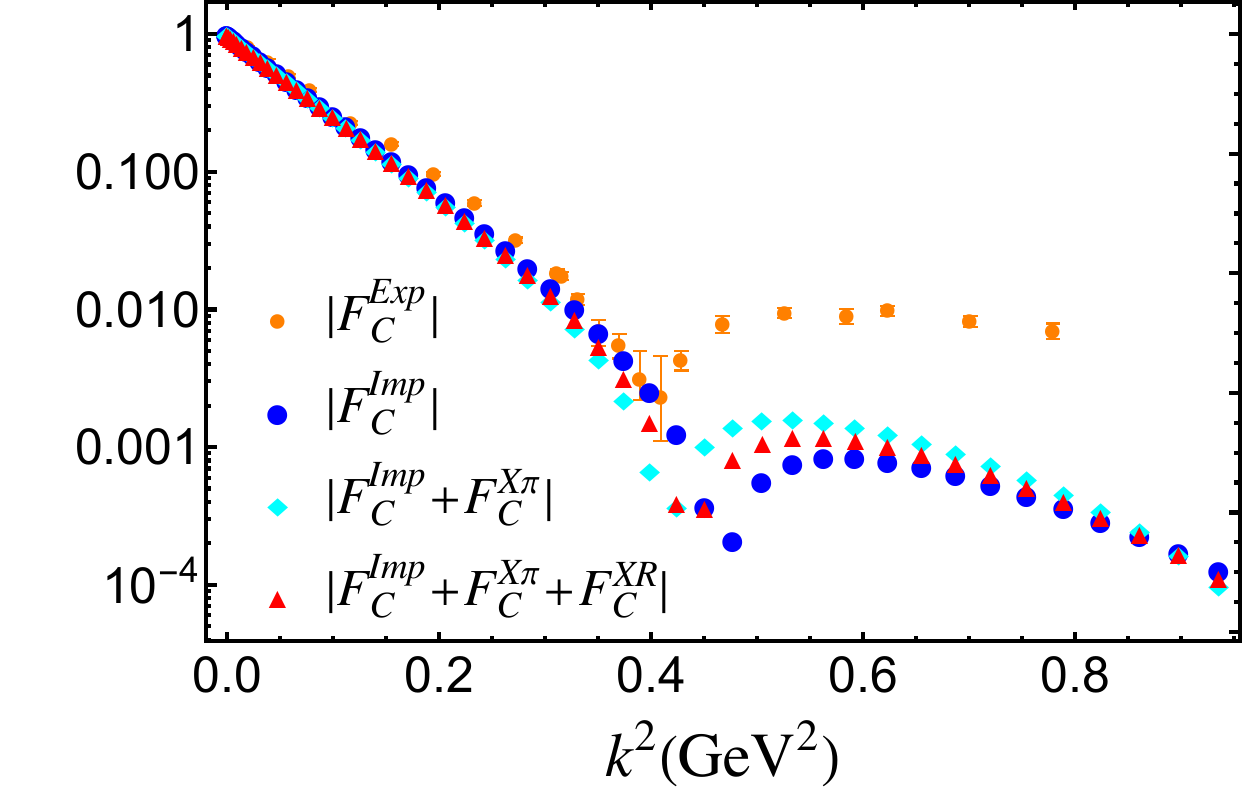}   
 \vspace{0pt}
\end{minipage}
\hfill
\begin{minipage}[b]{.45\linewidth}   
\hspace*{-0.55cm} \includegraphics[width=1.1\textwidth, height=5.5cm]{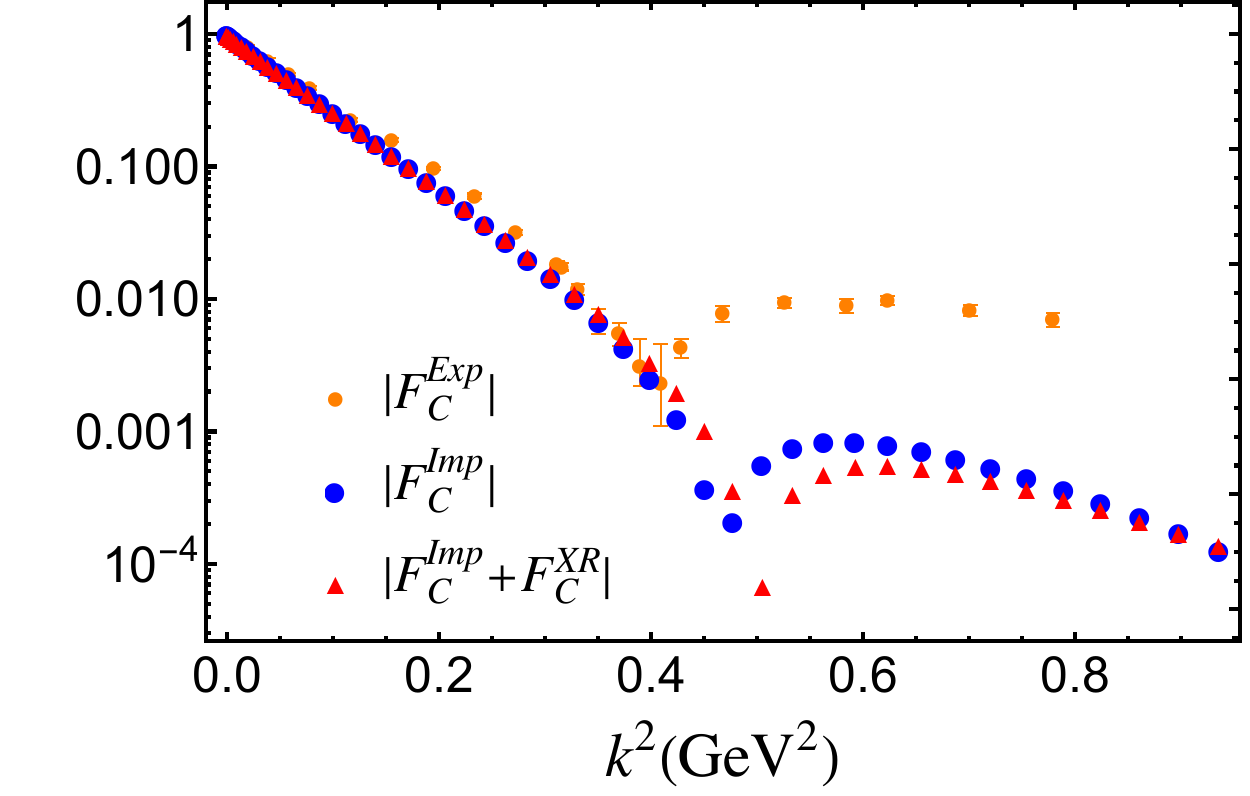} 
   \vspace{0pt}
\end{minipage}  
\caption{Charge form factor of Helium-4 with the PS coupling(left) and PV coupling (right) using the K-Harmonics method. The subscript ``imp" refers to the impulse approximation, while the subscripts  ``$X\pi$" and ``XR" refer to the exchange contributions from 
the pair diagram shown in~Fig.~\ref{fig:NR_diagrams}a and recoil contribution shown in~Fig.~\ref{fig:NR_diagrams}d, respectively. ``Exp" are the experimental results for the charge form factor in~\cite{Garcon:1993vm}.}
\label{fig:ex_resultCKH}
\end{figure*}

\begin{figure*}[htbp] 
\begin{minipage}[b]{.45\linewidth}
\hspace*{-0.3cm}\includegraphics[width=1.1\textwidth, height=5.5cm]{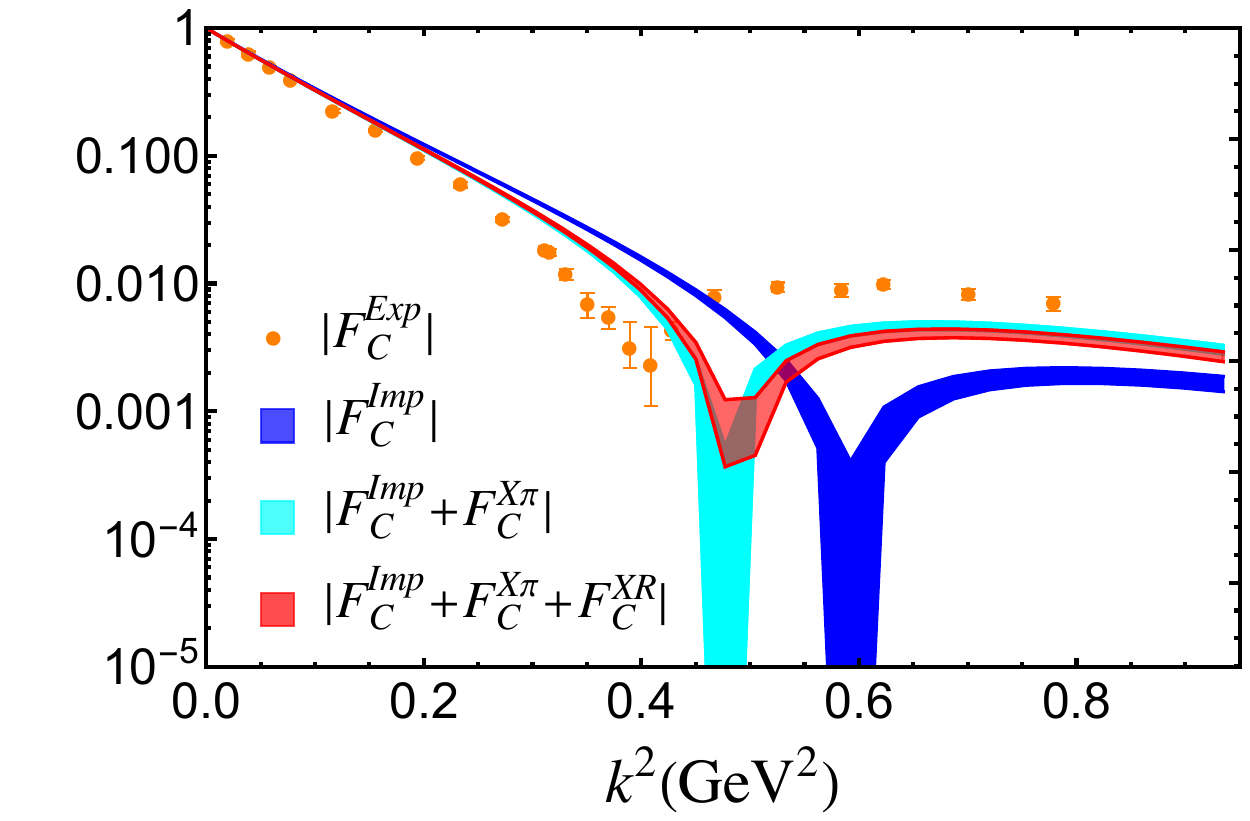}   
 \vspace{0pt}
\end{minipage}
\hfill
\begin{minipage}[b]{.45\linewidth}   
\hspace*{-0.55cm} \includegraphics[width=1.1\textwidth, height=5.5cm]{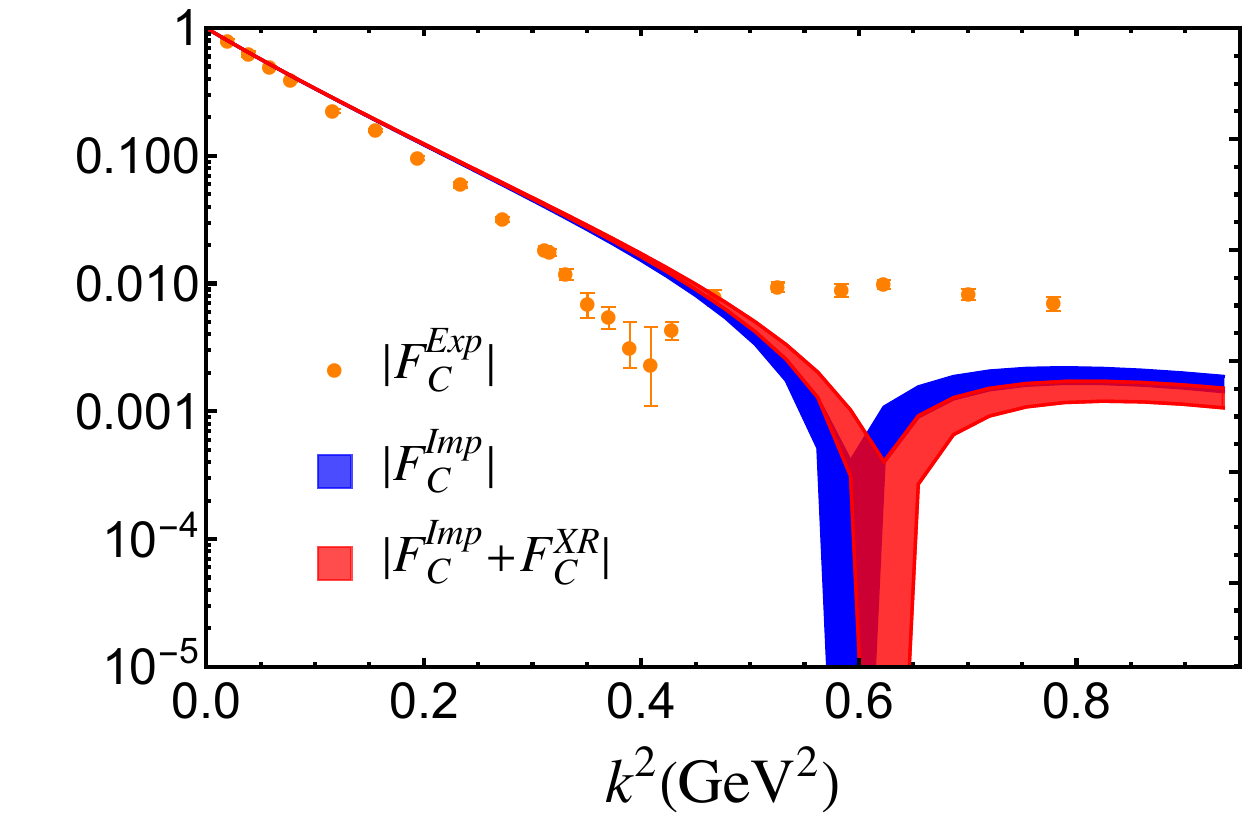} 
   \vspace{0pt}
\end{minipage}  
\\[-0.4cm]
\caption{The charge form factor of Helium-4 with PS coupling (left) and PV coupling (right) using the Monte Carlo method. The bands represent the statistic uncertainty. The orange data points are the experimental results in~\cite{Garcon:1993vm}. }
\label{fig:ex_resultCMC} 
\end{figure*}

\section{Conclusions}
\label{SECVI}
We have analyzed the exchange contributions to the GFFs $A,D$ for Helium-4. For that, we used a wavefunction that is maximally symmetric in the hyper-radius following from the K-harmonic method, with no D-wave admixture for simplicity. The results for the  form factor $A^H$ are overall consistent with 
the one we recently reported using the impulse approximation in~\cite{He:2023ogg}, whether we use the pion-nucleon PS or PV couplings. The diffractive minima of the form factor $D^H$ shift leftward when the  exchange currents are added.

For comparison, we have also used a more realistic  wavefunction following from the  Argonne $v14$ potential, with D-wave admixture. 
This approach is numerically intensive, and relies on Monte-Carlo
sampling to generate the GFFs, with inherent statistical uncertainties
that can be improved with longer runs. In this case, the exchange contributions were found to be overall consistent with the impulse approximation within statistics, when the pion-nucleon PS coupling is used. When the PV coupling is used,  the  exchange contributions are significant.

Our analysis of the charge form factor  for Helium-4 carried for comparison in Appendix~\ref{APPII}. The exchange contributions from the PS coupling  to the Helium-4 charge form factor, are larger in comparison to the impulse approximation when using the Argonne $v14$ potential, significantly improving the agreement with experiment, while the PV coupling  contributions are smaller and in agreement with the impulse approximation. For both cases, the diffractive GFF minima
following from the impulse approximation, shift rightward in comparison to the K-harmonic result.

Overall, most of the  exchange contributions are small in the range $k\leq \frac 12 m_N$, leaving unchanged the mass radii. While they are separately 
sizable at higher momenta, fortunately their sum total is small,  with
the exception of the GFFs with PV coupling used with the Argonne $v14$
potential. Since the Helium charge form factor is better described by
the PS coupling using the Argonne $v14$ potential, we expect the PS
results for the GFFs with this potential to be more reliable.

The nucleon GFFs have been recently probed at JLab~\cite{GlueX:2019mkq,Duran:2022xag}, through near threshold photoproduction of charmonium. It would be  useful to extend these experiments to Helium-4, for a better understanding of the role of the exchange currents in understanding the mass distribution in light nuclei,
with a particular interest in extracting the pion GFF.

\vskip 0.5cm
{\noindent\bf Acknowledgements}

\noindent 
We thank Zein-Eddine Meziani and Robert Wiringa  for discussions.
FH is supported by the National Science Foundation under CAREER Award PHY- 1847893. IZ is supported by the Office of Science, U.S. Department of Energy under Contract  No. DE-FG-88ER40388. This research is also supported in part within the framework of the Quark-Gluon Tomography (QGT) Topical Collaboration, under contract no. DE-SC0023646.

\appendix

\section{Trial wave functions}
\label{sec:wfv14}
In table~\ref{tab:wfslov}, we list the numerical values of three wave-functions $f_{0,1}$, $f_{1,0}$, $f_{t,0}$ solution to the coupled Schrodinger equations~(\ref{eq:seq}) in the regime $r<10$ fm. 
In the regime $r>10$ fm, we use the asymptotic expressions in~\cite{Wiringa:1991kp},

\bea
f_{0,1}(r>10fm)&=&1.88\left(\frac{e^{-0.850961 r}}{r}\right)^{\frac 13}, \nonumber\\
f_{1,0}(r>10fm)&=&1.75 \left(\frac{e^{-0.912554 r}}{r}\right)^{\frac 13}, \nonumber\\
f_{t,0}(r>10fm)&=&0.0665 \left(1-e^{-\frac{r^2}{4}}\right) \nonumber\\
&\times&\left(\frac{3.6025}{r^2}+\frac{3.28748}{r}+1\right) 
 \nonumber\\
&\times&\left(\frac{e^{-0.912554 r}}{r}\right)^{\frac 13}.
\eea

\section{Electric form factor of Helium-4}
\label{APPII}
In this appendix, we summarise  the results for the charge form factor for Helium-4 using the K-Harmonic method, and the trial state using the Argonne $v14$ potential.  More specifically, the leading non-relativistic form for 
the isoscalar electric operators for the pair diagram with PS and PV couplings are
\bea
\label{eq:J0pair}
 J_{X\pi,PS}^{0}(k)&=&G_M^S\frac{g_{\pi N}^2}{(2m_N)^2}\tau_1\cdot\tau_2\frac{\vec{\sigma}_1.\vec{k}\vec{\sigma}_2.\vec{q_2}}{m_N w_{q_2}^2}.\nonumber\\ 
&+&(1\leftrightarrow 2)+{\cal O}\bigg(\frac {g_{\pi N}^2}{m_N^4}\bigg),
\nonumber\\
J_{X\pi,PV}^{0}(k)&=&{\cal O}\bigg(\frac{g_{\pi N}^2}{m_N^4}\bigg).
\eea
The operator for the recoil correction for both PS and PV is
\bea
\label{eq:J0R}
J^{0}_{XR}
&=&-\frac{g_{\pi N}^2}{(2m_N)^2}\tau_1\cdot\tau_2G_E^S\frac{q_2.k \sigma_1.q_2 \sigma_2.q_2}{2 m_N w_{q_2}^4}\nonumber\\ 
&+&(1\leftrightarrow 2)+{\cal O}\bigg(\frac {g_{\pi N}^2}{m_N^4}\bigg),
\eea
where $G_M^S$ and $G_M^S$ represent the iso-singlet combination
of nucleon electric and magnetic form factors. The seagull diagram and pion exchange diagrams only contribute to the isospin vector form factor, not discussed here. The results of electric form factors can be easily obtained since the operator $J_{X\pi,PS}^{0}(k)$ and $J^{0}_{XR}$ are quite similar as $T_{X\pi,PS}^{00}(k)$ and $T^{00}_{XR}$ defined in Eq.~(\ref{eq:Tmunupair_PS00}) and Eq.~(\ref{eq:TmunuR00}).

In Fig.~\ref{fig:ex_resultCKH} we show the results for Helium-4 charge form factor with the PS pion-nucleon coupling (left) and PV pion-nucleon coupling (right) using the K-Harmonics method with no D-wave admixture.  The labels are  ``$Imp$" (impulse approximation),   ``$X\pi$" 
(pair correction), ``$XR$" (recoil correction) and ``$Exp$" (experiment) from~\cite{Garcon:1993vm}. For comparison, we show 
in Fig.~\ref{fig:ex_resultCMC} the results of the Monte-Carlo simulation using the wavefunction generated from the Argonne $v14$ potential with D-wave admixture, with the same labelings and color coding. The colored-spreads reflect on the statistical uncertainties.
In the latter, the PS coupling yields a charge form factor that is in reasonable agreement with the data.

\bibliography{refdeuteron}
\end{document}